\documentclass[acmsmall]{acmart}

\usepackage[english]{babel}

\usepackage{amsthm,amsfonts}
\usepackage{algorithmic}
\usepackage[ruled, noline, linesnumbered]{algorithm2e}
\usepackage{xcolor}

\usepackage{array}
\usepackage{multirow}
\usepackage[shortlabels]{enumitem}
\usepackage{tikz}

\usepackage[font=footnotesize]{subcaption}

\usepackage{textcomp}
\AtBeginDocument{
  \providecommand\BibTeX{{
    \normalfont B\kern-0.5em{\scshape i\kern-0.25em b}\kern-0.8em\TeX}}}

\usepackage{lipsum}


\theoremstyle{plain}
\newtheorem{theorem}{Theorem}

\theoremstyle{definition}

\def\({\left(}
\def\){\right)}
\def\[{\left[}
\def\]{\right]}
\def\{{\left\lbrace}
\def\}{\right\rbrace}

\def\LDA{{\rm LDA}}
\def\AoI{{\rm AoI}}

\def\st{{\rm s.t.}}
\def\OptConsSep{&&\quad}

\def\IFIL{IFIL\xspace}
\def\LIFOW{LIFO-W\xspace}

\def\fig#1{Fig.~\ref{fig:#1}}
\def\subfignum#1#2{(\subref{subfig:#1-#2})}
\def\subfig#1#2{Fig.~\ref{fig:#1}\subfignum{#1}{#2}}
\def\sec#1{Section~\ref{sec:#1}}
\def\thm#1{Theorem~\ref{thm:#1}}
\def\eqn#1{\eqref{eqn:#1}}

\renewcommand{\paragraph}[1]{\vspace*{0.3\baselineskip}\noindent {\bf #1:} }

\newcommand{\OptMax}[2]{
\begin{alignat}{2}
\max\ &\ #1 \nonumber \\
\st\ #2 
\end{alignat}
}

\newcommand\OptCons[3]{
&\ #1 
\ifx\\#2\\ \else \OptConsSep #2 \fi%
\ifx\\#3\\ \nonumber \else \label{eqn:#3} \fi%
}

\title{\ifcsname TitlePrefix\endcsname%
\csname TitlePrefix\endcsname%
\fi%
Trading Throughput for Freshness: Freshness-Aware Traffic Engineering and In-Network Freshness Control}

\setcopyright{none}

\copyrightyear{2021}
\acmYear{2021}
\acmDOI{}

\author{Shih-Hao Tseng}
\orcid{0000-0003-2376-9333}
\affiliation{
  \institution{California Institute of Technology}
  \streetaddress{1200 E. California Blvd.}
  \city{Pasadena} 
  \state{California} 
  \postcode{91125}
}
\email{shtseng@caltech.edu}
\author{SooJean Han}
\affiliation{
  \institution{California Institute of Technology}
  \streetaddress{1200 E. California Blvd.}
  \city{Pasadena} 
  \state{California} 
  \postcode{91125}
}
\email{soojean@caltech.edu}
\author{Adam Wierman}
\affiliation{
  \institution{California Institute of Technology}
  \streetaddress{1200 E. California Blvd.}
  \city{Pasadena}
  \state{California}
  \postcode{91125}
}
\email{adamw@caltech.edu}

\renewcommand{\shortauthors}{Tseng et al.}

\begin{document}

\def\ResultRatea{0.125}
\def\ResultRateb{5}
\def\ResultRatec{49}
\def\ResultRated{71}
\def\ResultRatee{10}
\def\ResultRatef{13}
\def\ResultRateg{7}
\def\ResultRateh{12}
\def\ResultRatei{5}
\begin{abstract}
In addition to traditional concerns such as throughput and latency, freshness is becoming increasingly important. To stay fresh, applications stream status updates among their components. Existing studies propose the metric age of information (AoI) to gauge the freshness and design systems to achieve low AoI. Despite active research in this area, existing results are not applicable to general wired networks for two reasons. First, they focus on wireless settings where AoI is mostly affected by interference and collision while queueing is more dominant in wired settings. Second, the legacy drop-adverse flows are not taken into account in the literature. Scheduling mixed flows with distinct performance objective is not yet addressed.

In this paper, we study wired networks shared by two classes of flows, aiming for high throughput and low AoI respectively, and achieve a good trade-off between their throughput and AoI. Our approach to the problem consists of two layers: freshness-aware traffic engineering (FATE) and in-network freshness control (IFC). FATE derives sending rate/update frequency for flows via optimization, and its solution is then enforced by IFC through efficient scheduling mechanisms at each outport of in-network nodes. We also present efficient Linux implementation of IFC and demonstrate the effectiveness of FATE/IFC through extensive emulations. Our results show that it is possible to trade a little throughput ($\ResultRateb \%$ lower) for much shorter AoI ($\ResultRatec$ to $\ResultRated\%$ shorter) compared to state-of-the-art traffic engineering.
\end{abstract}

\maketitle

\section{Introduction}\label{sec:introduction}

As we step into the era of the Internet of Things (IoT), applications increasingly depend on the network to not only carry high throughput but also deliver the information when it is \emph{fresh} so that they can synchronize their geo-distributed components. Freshness is critical for emerging streaming and IoT applications. On cloud gaming platforms such as Google's Stadia \cite{Google-Stadia-Cloud-Gaming}, games are streamed from the servers to the players and freshness determines players' gaming experience; live-streaming services such as Facebook Live \cite{Facebook-Live} and YouTube TV \cite{YouTube-TV} strive to provide fresh content to their audiences. The rapid development of smart city \cite{phadke1993synchronized,terzija2011wide,celli2014dms,primadianto2016review}, connected vehicles \cite{gandhi2007pedestrian, talak2016speed, biomo2014routing, hoel2019combining, talak2019optimizing}, and IoT technologies also introduces a diversity of safety-critical applications that require consistently available fresh and up-to-date information. Given these applications, freshness as an end-to-end performance metric is outpacing more traditional measures like flow-level throughput and packet-level latency, and freshness-centric flows are expected to occupy a significant share of network traffic in the near future~\cite{cisco2019global}.

\begin{figure}
\centering
\includegraphics[scale=1]{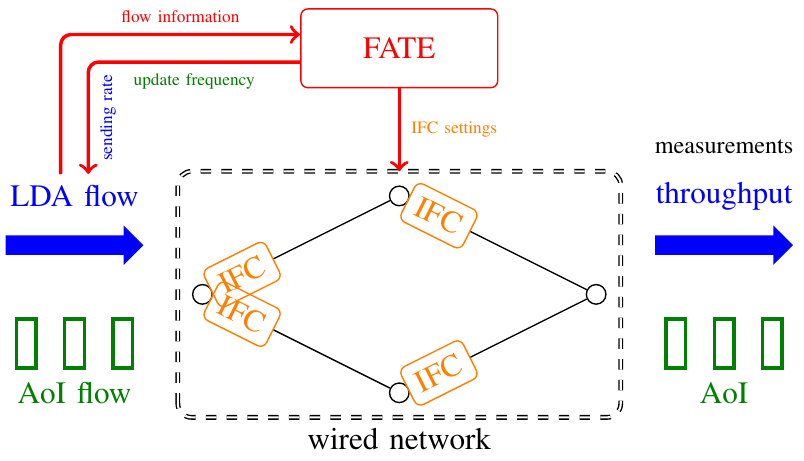}
\caption{We study a wired network with mixed legacy drop-adverse (LDA) flows and freshness-centric AoI flows, where LDA flows aim for high throughput while AoI flows aim for low age of information (AoI). We propose a two-layer system to achieve a good trade-off between the throughput and AoI. The freshness-aware traffic engineering (FATE) at the higher layer calculates the sending rate/update frequency for LDA/AoI flows, and the in-network freshness control (IFC) at the lower layer schedules the packets at the outports of the network nodes to enforce the FATE solution.
}
\label{fig:system-overview}
\end{figure}

To evaluate freshness, \emph{Age of Information (AoI)} has been proposed as a metric to gauge how synchronized the information source and its recipient are \cite{kaul2011minimizing,kaul2012real}. Multiple studies design information update systems to minimize AoI through various techniques, including update frequency control \cite{kaul2012real,yates2019age,sun2016update,sun2017update}, 
active queue management \cite{costa2014age,pappas2015age,costa2016age,
bedewy2016optimizing,bedewy2017age,bedewy2019minimizing}, and sending slots scheduling at base stations/interfered channels \cite{kadota2018optimizing,kadota2018scheduling,kadota2019scheduling,
talak2018scheduling,talak2019optimizing,talak2020improving}. We refer the reader to \cite{yates2019age,lou2021boosting} for surveys.

Despite enormous efforts on AoI minimization, existing proposals are not directly applicable to general wired networks for two reasons. First, most work in the literature focuses on wireless environments. 
Several papers focus on simple queueing system, single-hop base station, or probabilistic modeled network scheduling \cite{kaul2012real, yates2012real,kam2013age,costa2014age, huang2015optimizing,pappas2015age, bedewy2016optimizing, costa2016age, sun2017update,kadota2018optimizing, kadota2018scheduling, kavitha2018controlling, lu2018age, talak2018scheduling, altman2019forever,bedewy2019minimizing, kadota2019scheduling,yates2019age}, and some recent attempts start investigating multi-hop scenarios 
and trajectory planning for graph exploration \cite{bedewy2017age, liu2018age,talak2019optimizing, tripathi2019age, lou2020aoi,lou2021boosting, tripathi2021age}. Wireless environment is inherently different from wired environment in that the major source of delay comes from interference and collision while queueing plays a central role in the wired environment.

Second, existing work does not take the legacy drop-adverse (LDA) flows into account. Existing work considers the networks with only AoI flows -- the status update flows that aim for low AoI. While in practice, AoI flows will share the network with already-existing LDA flows that aim for either high throughput or low latency.
Some recent papers study the AoI flows with minimum throughput requirements (``dual intent'') \cite{kadota2018optimizing,kadota2019scheduling,lou2020aoi,lou2021boosting}. 
However, that setting is still different from mixed LDA and AoI flows, which present different opportunities and challenges as the flows have distinct, instead of coherent, interests. 
Especially, dropping LDA packets causes performance degradation, and hence the traditional network design avoids packet drops. In contrast, dropping AoI packets could potentially improve the freshness of the AoI flows \cite{bedewy2017age,kavitha2018controlling,bedewy2019minimizing,yates2019age}. As a result, it is possible to maximize the throughput for LDA flows while minimizing AoI for AoI flows by handling LDA and AoI packets differently. How to efficiently serve these two intents by packet management is not yet addressed in the literature. 

\paragraph{Contributions and Organization}
In this work, we study wired networks with mixed LDA and AoI flows as shown in \fig{system-overview}, where the two classes of flows aim for distinct performance goals -- high throughput and low AoI, respectively. We aim to achieve different trade-offs between LDA flows' throughput and AoI flows' AoI through our two-layer network design. The higher layer, the freshness-aware traffic engineering (FATE), optimizes sending rate/update frequency of the LDA/AoI flows. FATE centers on the LDA-AoI Coscheduling problem (LAC), which trades off LDA flows' throughput for AoI flows' AoI. LAC leverages inverted update frequency as a proxy for AoI, and we justify this choice by showing that a zero-queueing schedule is NP-hard to obtain and providing an upper bound on the AoI.

Given a FATE solution, the lower layer, the in-network freshness control (IFC), maintains the update frequency across the network through efficient per-port queueing management -- AoI-Aware Queueing (AAQ). We propose two different in-network scheduling policies for AAQ: size-driven multiplexing (SDM) and time-division multiplexing (TDM). We provide simple and effective implementations, and we demonstrate their low overhead and good performance through experiments.

We organize the paper as follows. We first briefly review the metric of AoI and discuss the techniques to balance throughput with freshness in \sec{background}. Then, in \sec{FATE}, we introduce FATE, formulate LAC, and analyze the factors that affect AoI to justify LAC. Given an LAC solution, we establish the IFC scheme through two policies SDM and TDM in \sec{IFC} and provide the efficient Linux implementation of IFC in \sec{implementation}. In \sec{evaluation}, we conduct experiments to examine the overhead of IFC, the effectiveness of FATE and IFC, and the combination of IFC with other queueing management methods. With related work reviewed in \sec{related-work}, we conclude the paper in \sec{conclusion}.
\section{Background and Motivation}\label{sec:background}
We start by introducing the age of information (AoI) metric for measuring freshness. Then, we explore the techniques that could balance throughput with freshness in a mixed LDA/AoI environment and discuss the corresponding challenges. 

\subsection{Freshness and Age of Information (AoI)}\label{sec:background-freshness-aoi}
\begin{figure}
\centering
\includegraphics{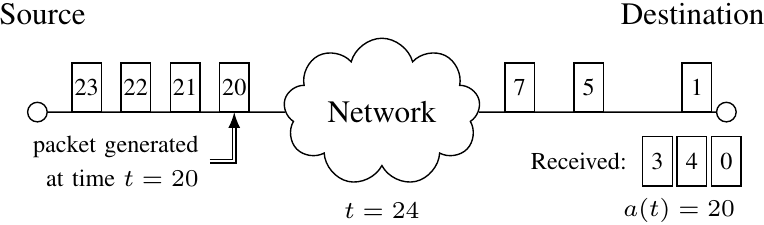}
\vspace*{-0.5\baselineskip}
\caption{The status update age $a(t)$ at time $t$ is the elapsed time since the freshest update received at the destination at time $t$ was generated at the source. In this example, the current time $t = 24$ and the generated time of each packet is marked accordingly.
Upon the arrival of the packet generated at time $t = 1$, the freshest update received at the destination was generated at $t = 4$, and hence the status update age is $a(t) = a(24) = 24 - 4 = 20$ at time $t = 24$. Notice that receiving the outdated (or ``out-of-ordered'') update (generated at $t = 1$) does not help reduce $a(24)$.}
\label{fig:aoi_definition}
\end{figure}

\begin{figure}
\centering
\includegraphics[scale=0.8]{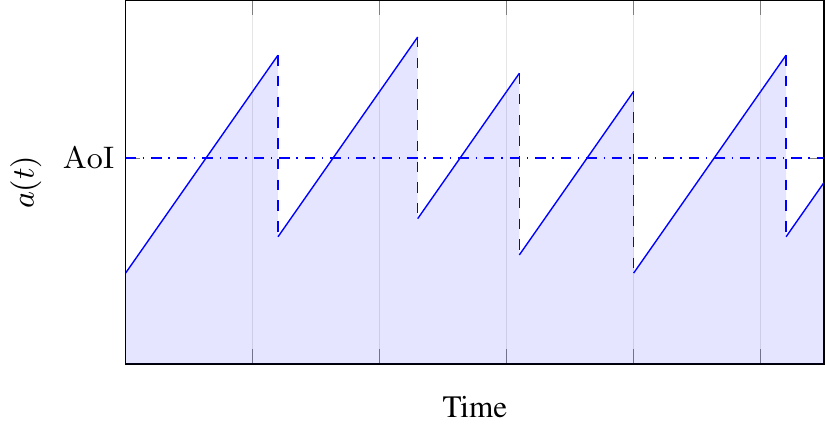}
\vspace*{-0.5\baselineskip}
\caption{Age of information (AoI) is defined as the long-term time-averaged status update age, which averages $a(t)$ (the shaded area) over time.}
\label{fig:long_term_time_averaged_aoi}
\end{figure}

A new metric that has recently emerged to quantify freshness is the \emph{Age of Information (AoI)} \cite{kaul2011minimizing}, which quantifies the degree of synchronization between the information source and its destinations. To synchronize the destinations, the information source streams status update packets through the network. The \emph{status update age} $a(t)$ is then defined as the elapsed time since the freshest update received at the destinations at time $t$, as illustrated in \fig{aoi_definition}.
$a(t)$ reduces upon the receipt of a new update, which results in the saw-like waveform in~\fig{long_term_time_averaged_aoi}.

Instead of directly referring to the fluctuating $a(t)$, we would quantify the degree of synchronization by the long-term time-averaged performance. Accordingly, we define AoI as the long-term time-averaged status update age, which is given by
\begin{align}
{\rm AoI} = \lim\limits_{T \to \infty} \frac{1}{T}\int_{0}^{T} a(t) \ dt.
\label{eqn:aoi_definition}
\end{align}
This is illustrated in \fig{long_term_time_averaged_aoi}. The existence of the above limit is typically assumed implicitly.

We remark that AoI/freshness is different from latency. Latency measures how fast a packet can be delivered from the source to the destination, while freshness evaluates how synchronized the source and the destination are. The former is a per-packet property, while the latter depicts an end-to-end characteristic.
Although shorter latency can potentially help achieve fresher information delivery, freshness does not require a fast or even intact transmission for all packets. For example, dropping a packet hurts the latency of transmitting the packet, but it might clear the path for fresher information to arrive at the destination and thus improve the AoI.

\subsection{Balancing Throughput with Freshness}\label{sec:background-tradeoff}

\begin{figure}
\centering
\subcaptionbox{An LDA flow and an AoI flow share a downstream link through a node with infinite-sized buffer. AoI grows indefinitely as old AoI packets are buffered.
\label{subfig:background_balancing-share}}{
\includegraphics[scale=0.8]{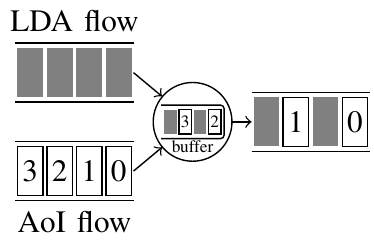}
}\hfill
\subcaptionbox{Naively prioritizing AoI over LDA causes throughput breakdown. Prioritizing LDA over AoI will work neither -- no update can be delivered. \label{subfig:background_balancing-prioritize_AoI}}{
\includegraphics[scale=0.8]{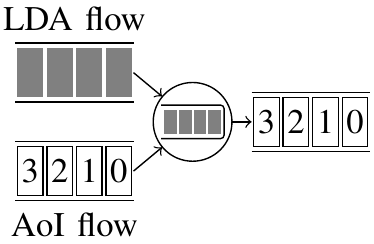}
}\hfill
\subcaptionbox{Halving the update frequency at the source avoids buffering AoI packets and achieves a steady short AoI.
\label{subfig:background_balancing-rate_control}}{
\includegraphics[scale=0.8]{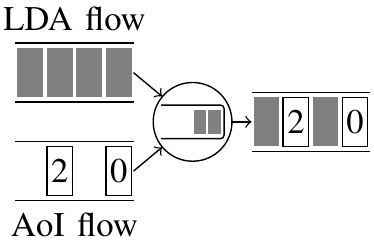}
}\hfill
\subcaptionbox{Alternatively, we can drop/replace outdated queued packets at the node for fresh packet delivery and short AoI. \label{subfig:background_balancing-IFC}}{
\includegraphics[scale=0.8]{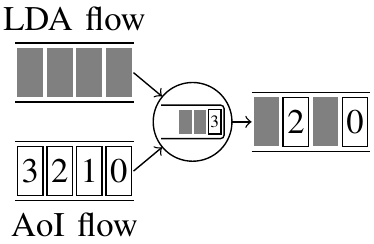}
}
\caption{Balancing the throughput of LDA flow with the AoI of the AoI flow requires techniques more sophisticated than simple prioritization. We could perform update frequency control or active packet drop to improve the AoI.
}
\label{fig:background_balancing}
\end{figure}

We refer to the status update flows that aim for freshness the \emph{AoI flows}. In addition to the new AoI flows, legacy network flows pursue high throughput/low latency. Since dropping packets hurts both throughput and latency, we call those legacy flows the \emph{legacy drop-adverse (LDA) flows}. Traditional networks are designed to serve LDA flows, and hence most network techniques center around packet-drop avoidance.

Classical network design may not be beneficial for the freshness-centric AoI flows. Consider the example shown in \subfig{background_balancing}{share}. Three links of unit capacity connect to a node with infinite-sized buffer in the middle. An LDA flow and an AoI flow traverse through links, join at the node, and share the same downstream link. Traditional network design might share the link equally, which causes the AoI packets buffered at the node and the end-to-end AoI will grow indefinitely. 

To avoid buffering AoI packets, one might want to prioritize the AoI flow over the LDA flow (\subfig{background_balancing}{prioritize_AoI}). However, that completely blocks the LDA flow and results in throughput breakdown. Therefore, to well balance throughput with AoI, we need more sophisticated approaches than naive prioritizing.

There are two techniques that can achieve a more balanced trade-off: update frequency control and active packet drop. As shown in \subfig{background_balancing}{rate_control}, we can halve the update frequency of the AoI flow. As such, no AoI packet will be buffered and we achieve a smaller end-to-end AoI. Alternatively, we can also buffer and drop/replace outdated AoI packets at the intermediate node as in \subfig{background_balancing}{IFC}. Accordingly, we always send out fresh packets rather than letting packets age in the queue.

To perform these techniques, we need to address a few challenges. First, how can we come up with the right update frequency for the AoI flows? Also, in a practical network setting, buffers are of finite sizes and we might also need to determine the sending rate of the LDA flows to avoid packet drops. In addition, how should we drop the packets? Especially, we need to discard outdated packets efficiently without incurring too much processing overhead. Otherwise, burdened network nodes would suffer slow forwarding and hurt freshness. In this paper, we deal with these issues through FATE, IFC, and efficient Linux implementation that are elaborated in the following sections.

\def\picscale{0.85}

\section{Freshness-Aware Traffic Engineering (FATE)}\label{sec:FATE}
In this section, we establish our freshness-aware traffic engineering (FATE). FATE models LDA flows as fluids of fixed sending rate and AoI flows as sequences of status updates with fixed update frequency. Given predetermined routes, FATE performs sending rate/update frequency control over LDA/AoI flows to optimize some objective. In particular, we propose the LDA-AoI Coscheduling (LAC) problem, which balances throughput and an approximated AoI. The approximation is then justified through our analysis of the composition of AoI, which shows that a zero-queueing schedule is NP-hard to verify/obtain and that AoI can be upper-bounded.

\subsection{Model}
\begin{figure}
\centering
\includegraphics[scale=1]{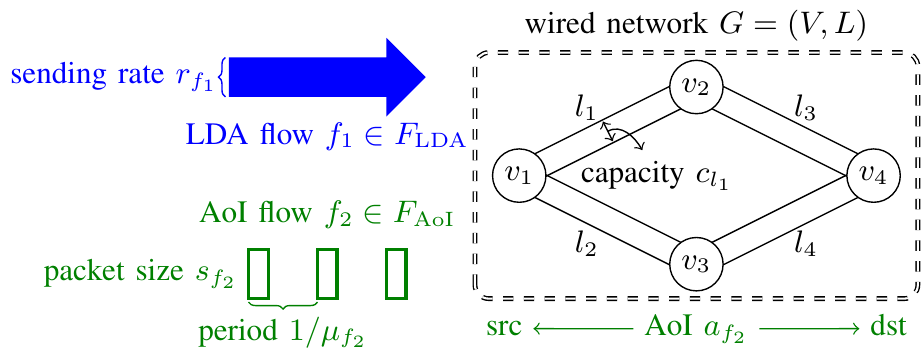}
\caption{We model the network by a directed graph $G = (V,L)$ shared by a set of LDA flows $F_\LDA$ and a set of AoI flows $F_\AoI$. Freshness-aware traffic engineering (FATE) optimizes over the sending rate $r_f$ for $f \in F_\LDA$ and the update frequency $\mu_f$ for $f \in F_\AoI$.}
\label{fig:FATE-model}
\end{figure}

We model the network by a directed graph $G = (V,L)$ where $V$ is the set of nodes and $L$ the set of links. Each link $l \in L$ has a non-negative capacity $c_l$. Two sets of flows $F_\LDA$ and $F_\AoI$ share the network, where $F_\LDA$ consists of LDA flows and $F_\AoI$ is the set of AoI flows. Each flow $f \in F_\LDA \cup F_\AoI$ sends through a predetermined simple path from some source node to some destination node. We write $l \in f$ if and only if link $l$ is on the path of flow $f$. 
We denote by $r_f$ the throughput (sending rate) of an LDA flow $f \in F_\LDA$ and by $a_f$ the AoI of an AoI flow $f \in F_\AoI$. To compute AoI, we model each AoI flow $f \in F_\AoI$ as a periodic status update sequence with update frequency (number of updates per unit time) $\mu_f$. Each status update is a sequence of packets with total size $s_f$. Without loss of generality, we model each status update as one AoI packet of size $s_f$ in the following context. \fig{FATE-model} provides an illustration of the notation defined here.

\subsection{LDA-AoI Coscheduling (LAC) Problem}
FATE computes the sending rate/update frequency for LDA/AoI flows by solving the LDA-AoI Coscheduling (LAC) problem. LAC aims to maximize the throughput of LDA flows while minimizing the AoI by finding the best update frequency for the AoI flows. To do so, LAC optimizes the following objective
\begin{align*}
\max \sum\limits_{f \in F_\LDA} r_f - \lambda \sum\limits_{f \in F_\AoI} \frac{1}{2\mu_f}
\tag{LAC}
\end{align*}
where $\lambda \geq 0$ is the trade-off factor. 

As shown in \subfig{background_balancing}{share}, congestion would hurt AoI as the old AoI packets are buffered. Therefore, LAC strives for congestion-free sending/status update via capacity constraints as follows. Let $S_\LDA(l)$ and $S_\AoI(l)$ be the total LDA/AoI traffic on the link $l$, i.e.,
\begin{align*}
S_\LDA(l) = \sum\limits_{f\in F_\LDA : l \in f} r_f,\quad\quad
S_\AoI(l) = \sum\limits_{f \in F_\AoI : l \in f} \mu_f s_f.
\end{align*}
Assuming no packet drop, the capacity constraint requires, for each $l \in L$,
\begin{align}
S_\LDA(l) + S_\AoI(l) \leq c_l.
\label{eqn:cons:capacity}
\end{align}

Together, we can formulate the LDA-AoI coscheduling (LAC) problem as
\OptMax{
\sum\limits_{f \in F_\LDA} r_f - \lambda \sum\limits_{f \in F_\AoI} \frac{1}{2\mu_f} 
}{
\OptCons{S_\LDA(l) + S_\AoI(l) \leq c_l}{\forall l \in L}{}\\ 
\OptCons{r_f \geq 0}{\forall f \in F_\LDA}{}\\
\OptCons{\mu_f \geq 0}{\forall f \in F_\AoI}{}
}
We remark that LAC can be efficiently solved in the following equivalent convex form by second order cone methods:
\OptMax{
\sum\limits_{f \in F_\LDA} r_f - \frac{\lambda}{2} \sum\limits_{f \in F_\AoI} h_f
\tag{Equivalent Convex LAC}
}{
\OptCons{S_\LDA(l) + S_\AoI(l) \leq c_l}{\forall l \in L}{}\\ 
\OptCons{r_f \geq 0}{\forall f \in F_\LDA}{}\\
\OptCons{\mu_f \geq 0}{\forall f \in F_\AoI}{}\\
\OptCons{h_f \mu_f \geq 1}{\forall f \in F_\AoI}{}
}

Notice that LAC minimizes AoI via minimizing $\frac{1}{2\mu_f}$, which is an approximation to the true AoI $a_f$. We motivate and justify such an approximation in the following subsection.

\subsection{AoI Factors and Upper Bound}
Instead of minimizing AoI $a_f$ directly, LAC minimizes the inverse of $\mu_f$ as a proxy. This choice is driven by the following two results. First, it is NP-hard to precisely determine AoI due to the queueing delay. Second, we can upper-bound AoI, and the varying term in the bound is proportional to the inverse of $\mu_f$.
We derive these results by providing a more detailed analysis of AoI herein, and the proofs for the theorems in this subsection are deferred to Appendices.

\begin{figure}
\centering
\includegraphics[scale=0.8]{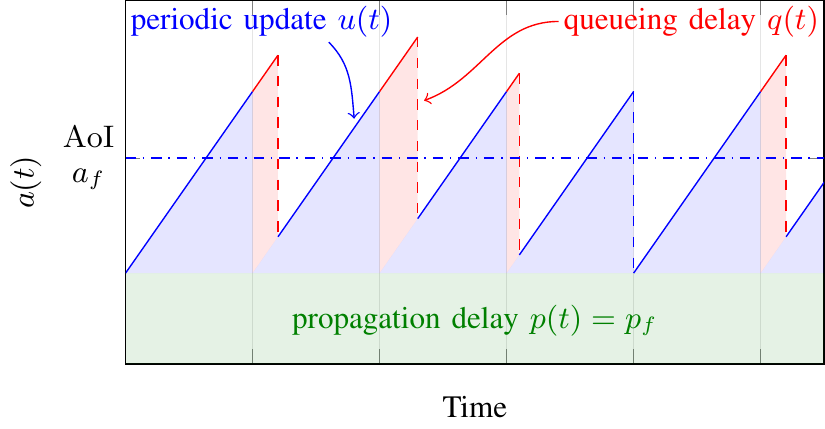}
\caption{Break down the cause of AoI: periodic update, propagation delay, and queueing delay.}
\label{fig:aoi-factors}
\end{figure}

As shown in \fig{aoi-factors}, the status update age $a(t)$ of a flow $f$ can be attributed to three factors: periodic update $u(t)$, propagation delay $p(t)$, and queueing delay $q(t)$, which satisfy
\begin{align*}
a(t) = u(t) + p(t) + q(t).
\end{align*}
The longer the update period, or the lower the update frequency $\mu_f$, of an AoI flow leads to long AoI. The propagation delay $p(t)$ accounts for the time one status update traverses through an empty network, which is a constant $p_f$ determined 
by the total link latency $d_f$ and the packet transmission delay, i.e.,
\begin{align*}
p_f = d_f + \sum\limits_{l : l \in f} \frac{s_f}{c_l}.
\end{align*}
The queueing delay $q(t)$ accounts for the time when an update is not forwarded right upon its arrival at an intermediate node and becomes stale while waiting in the queue. According to \eqn{aoi_definition}, the AoI $a_f$ of the flow $f$ is
\begin{align*}
a_f =&\ \lim\limits_{T \to \infty} \frac{1}{T}\int_{0}^{T} a(t) \ dt
= \lim\limits_{T \to \infty} \frac{1}{T}\int_{0}^{T} u(t) + p(t) + q(t) \ dt
= \frac{1}{2\mu_f} + p_f + q_f
\end{align*}
where $q_f$ is the long-term time-averaged $q(t)$. We slightly abuse the terminology to also refer to $q_f$ as the queueing delay.

The most uncertain term in the LAC problem is the queueing delay $q_f$. There are three scenarios under which an update suffers queueing delay -- waiting for LDA flows, previous updates, or updates from other AoI flows. To tackle these scenarios, we can prioritize AoI updates to avoid AoI flows being blocked by LDA flows. Also, if an update needs to wait for previous updates from the same flow, it suggests that the update frequency of the flow is higher than the forwarding link capacity. We should reduce the update frequency and discard old queued updates when a new update arrives.

\begin{figure}
\centering
\subcaptionbox{Status updates from different AoI flows overlap with each other and suffer queueing delay.\label{subfig:aoi-interleaving-need-interleave}}[0.48\columnwidth]{
\includegraphics[,scale=\picscale]{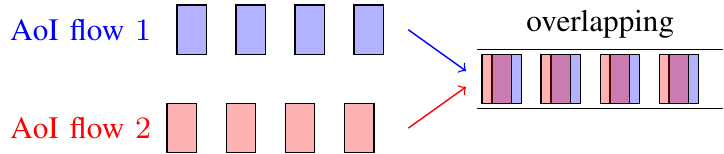}
}\hfill
\subcaptionbox{In this case, it is possible to interleave the AoI flows to avoid overlapping and queueing delay.\label{subfig:aoi-interleaving-can-interleave}}[0.48\columnwidth]{
\includegraphics[,scale=\picscale]{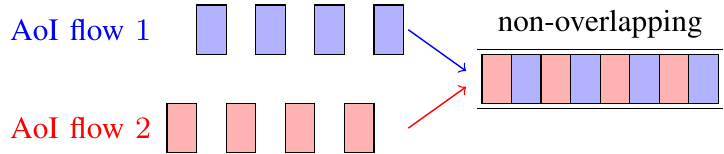}
}
\caption{Carefully interleaving the AoI flows can potentially mitigate queueing delay in some cases.
}
\label{fig:aoi-interleaving}
\end{figure}

\begin{figure}
\subcaptionbox{Queueing delay due to update frequency mismatch.\label{subfig:aoi-cannot-interleave-frequency-mismatch}}[\columnwidth]{
\includegraphics[,scale=\picscale]{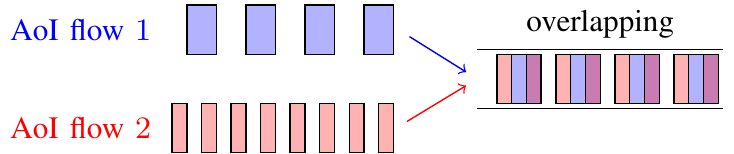}
}\\
\subcaptionbox{Queueing delay due to asynchronous arrival.\label{subfig:aoi-cannot-interleave-asynchrony}}[\columnwidth]{
\includegraphics[,scale=\picscale]{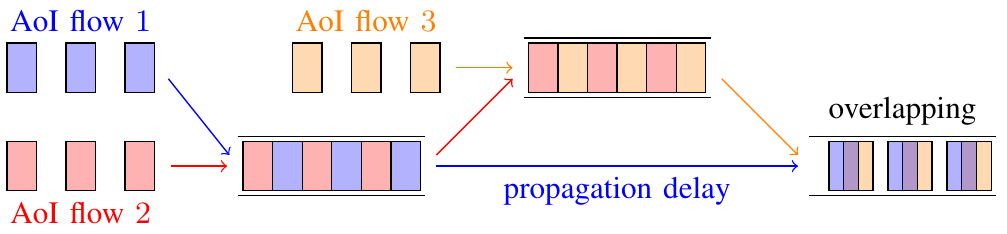}
}
\caption{It is not always possible to completely avoid queueing delay by interleaving the flows even though the links have enough capacities.}
\label{fig:aoi-cannot-interleave}
\end{figure}

The trickiest scenario is when AoI packets from different flows wait for one another. Though in some cases, such as in \subfig{aoi-interleaving}{need-interleave} and \subfignum{aoi-interleaving}{can-interleave}, we can carefully interleave the AoI flows to reduce the queueing delay, it is not always the case. Even if we ensure link capacity is enough to accommodate the updates in the long run through condition \eqn{cons:capacity}, the arrival timing could still cause transient congestion. For example, in \subfig{aoi-cannot-interleave}{frequency-mismatch}, both AoI flows occupy half of the link capacity. Since flow $2$ updates twice more frequent than flow $1$, it is not possible to interleave them without congestion. Another example in \subfig{aoi-cannot-interleave}{asynchrony} shows three AoI flows sharing different links. We can interleave flow $1$ and $3$ carefully to ensure they do not overlap with flow $2$. However, when they meet at the rightmost link, they inevitably collide and cause queueing delay. 

In fact, even when a zero-queueing-delay interleaving schedule is possible, obtaining such a schedule is NP-hard by the following theorem. 
\begin{theorem}\label{thm:zero-queueing-delay}
Given a set of AoI flows $F_\AoI$ with update frequencies $\mu_f$, status size $s_f$, and link capacities $c_l$ for all $f \in F_\AoI$ and $l \in L$, it is NP-hard to determine if there exists a zero-queueing-delay interleaving schedule.
\end{theorem}

While it is NP-hard to determine a zero-queueing-delay schedule, it is possible to upper-bound AoI (and hence upper-bound the queueing delay) for LAC solutions by the following theorem.
\begin{theorem}\label{thm:bounded-queueing-delay}
Given a feasible solution to LAC, there exists a scheduling policy such that 
\begin{align*}
a_f \leq \frac{1 + 2|f|}{2\mu_f} + d_f
\end{align*}
for all $f \in F_\AoI, \mu_f > 0$, where $|f| = | \{ l : l \in f \}|$ is the number of links on the path taken by $f$.
\end{theorem}
Since $1 + 2|f|$ and $d_f$ are both constants, we know that the varying part of this AoI upper bound is proportional to $\frac{1}{2 \mu_f}$, which is hence chosen as the proxy for AoI minimization in LAC.

\section{In-Network Freshness Control (IFC)}\label{sec:IFC}

Once we solve LAC, we can assign the sending rate and update frequency accordingly and the network should be congestion-free in the long run thanks to the condition \eqn{cons:capacity}. However, long-term freedom from congestion does not imply that the AoI will achieve its lowest value. 
For instance, the sending rate and update frequency might not be precisely applied at the source. Or, after going through multiple hops, queueing and stochastic processing delay could cause the update frequency of an AoI packet series to drift from the assigned value. 
Therefore, our mission then becomes enforcing the FATE solution into the network through in-network freshness control (IFC). We propose a new per-port queueing discipline -- AoI-aware queueing (AAQ) -- to do so. Below we present the design of AAQ and its core component, the scheduler.

\subsection{AoI-Aware Queueing (AAQ)}

\begin{figure}
\centering
\includegraphics{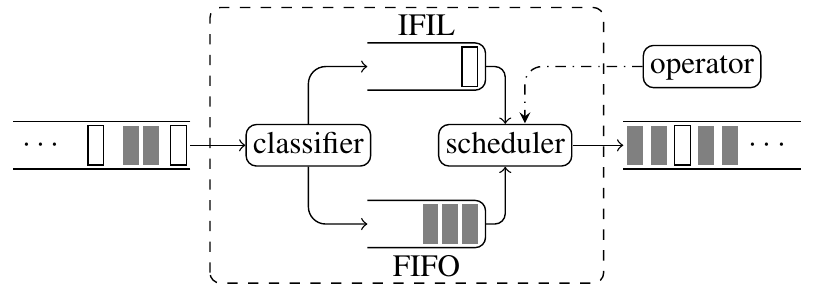}
\caption{To enforce an LAC solution, we introduce AoI-aware queueing (AAQ) at each out port to sort the packets before transmitting them on a link.}
\label{fig:aaq_overview}
\end{figure}

To start with, we make the following four simple observations:
\begin{enumerate}[<1>]
\item Work conservation: When $S_\LDA(l) + S_\AoI(l) < c_l$, we should increase $r_f$ and $\mu_f$ whenever possible.\label{observation-work-conservation}
\item Scheduler: We need a scheduler to allocate link capacities $c_l$ to LDA and AoI flows according to $S_\LDA(l)$ and $S_\AoI(l)$ in \eqn{cons:capacity}.\label{observation-scheduler}
\item Self queueing delay avoidance: We should replace the old AoI packet by the new one from the same AoI flow. \label{observation-self-queueing}
\item Cross queueing delay mitigation: AoI flows should be served in a FIFO manner.\label{observation-cross-queueing}
\end{enumerate}

Based on these observations, we propose a per-port queueing policy, the \emph{AoI-Aware Queueing (AAQ)}, to enforce an LAC solution. The basic idea of AAQ is to manage the output queue for each link in accordance with \eqn{cons:capacity}. The architecture of AAQ is shown in \fig{aaq_overview}.

For the traffic arrives at an outport, which is a mixture of LDA and AoI packets, AAQ sorts them into two sub-queues for LDA and AoI respectively by a classifier. Two sub-queues are governed by different queueing policies. AAQ chooses some queueing policy to maximize the throughput for the LDA sub-queue. For the AoI sub-queue, we propose the \emph{inter-flow FIFO, intra-flow \LIFOW (\IFIL)} based on our observations.

\IFIL, as its name suggests, enqueue different flows in FIFO manner, while the packets from the same flow follows \LIFOW policy: Last-arriving packet preempts existing waiting packet in the queue (if any). Operationally, every incoming AoI packet is enqueued at the back of the queue (observation \ref{observation-cross-queueing}). However, if there exists an AoI packet from the same flow in the queue already, the newly arrived packet preempts (replaces) the existing one (observation \ref{observation-self-queueing}).
In brief, \IFIL schedules different AoI flows in a FIFO manner, which avoids flow starvation, while keeping only the last arrived packet of each flow in the queue. This design also ensures that the number of status update packets buffered at each \IFIL queue is at most $|F_{\AoI}|$, since only the latest arriving packet is kept per AoI flow.

After the two sub-queues, AAQ installs a scheduler to reassemble the packets to the outgoing link (observation \ref{observation-scheduler}), which we will further elaborate on in the next subsection.

\subsection{Scheduler and Scheduling Policies}
A core component of AAQ is the scheduler that interleaves the LDA and AoI packets according to \eqn{cons:capacity} (observation \ref{observation-scheduler}) while conserving work (observation \ref{observation-work-conservation}). 
To do so, whenever the link is ready to send, the scheduler first check the occupancy of LDA and AoI subqueues. If only one of them carries packets, the scheduler dequeues packets to send from the non-empty queue. When both the subqueues are occupied, the scheduler decides which subqueue to dequeue from according to a scheduling policy that enforces $S_\LDA(l)$ and $S_\AoI(l)$. Below we discuss the design of such a policy.

To begin with, we define \emph{AoI ratio $\gamma_l$ of link $l$} to be the share of total traffic on link $l$ that is from AoI flows, i.e.,
\begin{align*}
\gamma_l = \frac{S_\AoI(l)}{S_\LDA(l) + S_\AoI(l)} 
\end{align*}
when $S_\LDA(l) + S_\AoI(l) > 0$; Otherwise, we set $\gamma_l = 1$. By definition, we have $\gamma_l \in [0,1]$ and 

when $\gamma_l = 1$, the link carries only AoI flows, while $\gamma_l = 0$ implies an LDA only link. In other words, $\gamma_l$ reflects the  LDA/AoI traffic distribution on link $l$.

To enforce the allocation $S_\LDA(l)$ and $S_\AoI(l)$ at each link $l$, we can keep track of the total LDA and AoI traffic on the link and ensure that the the ratio between them is equal to $1 - \gamma_l$ to $\gamma_l$ in the long run. Alternatively, given a fixed link capacity $c_l$, we can also ensure that the time spent on sending LDA/AoI traffic is proportional to $1 - \gamma_l$ and $\gamma_l$ correspondingly. These two options lead to the following \emph{size-driven multiplexing (SDM)} and \emph{time-division multiplexing (TDM)} policies. 

\paragraph{Size-Driven Multiplexing (SDM)} We maintain a budget variable $b_l$, initialized as $0$, for each link $l$. Whenever a LDA packet of size $s$ is sent over link $l$, we increment $b_l$ by $\gamma_l s$. Instead, when an AoI status update of size $s$ is sent, we decrement $b_l$ by $(1-\gamma_l) s$. The scheduling policy is then when $b_l > 0$, we prioritize AoI subqueue. Otherwise, we prioritize LDA subqueue. In the long run, SDM ensures the total LDA/AoI traffic sent through link $l$ achieves the desired AoI ratio $\gamma_l$.

\paragraph{Time-Division Multiplexing (TDM)} Given some fixed time frame $T$, we partition it into two sub-time frames $T_\LDA$ and $T_\AoI$ where
\begin{align*}
T_\LDA = (1-\gamma_l) T = \frac{S_\LDA}{S_\AoI + S_\LDA} T, \quad\quad 
T_\AoI = \gamma_l T = \frac{S_\AoI}{S_\AoI + S_\LDA} T.
\end{align*}
We take turns to prioritize LDA/AoI subqueues according to $T_\LDA$ and $T_\AoI$: We prioritize LDA subqueue for $T_\LDA$ then prioritize AoI subqueue for $T_\AoI$. By the end of each turn, if a packet/status update is still under transmission, we compensate the overtime proportionally to the other subqueue. As such, the ratio of the sending time for LDA/AoI traffic through link $l$ approaches $\frac{1-\gamma_l}{\gamma_l}$ in the long run. Correspondingly, the AoI ratio approaches to $\gamma_l$.

\begin{figure}
\centering
\subcaptionbox{Size-driven multiplexing (SDM)}{
\includegraphics[scale=0.9]{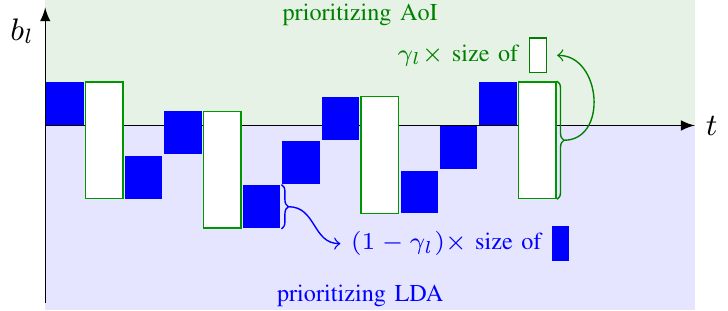}
}\hfill
\subcaptionbox{Time-division multiplexing (TDM)}{
\includegraphics[scale=0.9]{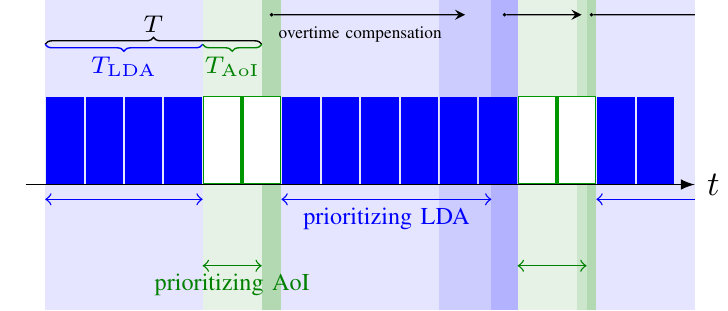}
}
\caption{An example showing the schedule fulfilled by SDM and TDM. The LDA packets are filled while the AoI packets are hollow. SDM maintains the variable $b_l$ to prioritize LDA/AoI subqueue accordingly, while TDM partitions the predetermined time frame $T$ to decide the prioritization periods. If a packet is still under transmission while the prioritization period ends, the overtime is compensated to the other subqueue.}
\label{fig:IFC_comparison_SDM_TDM}
\end{figure}

We use the example in \fig{IFC_comparison_SDM_TDM} to compare SDM with TDM. 
As shown in the figure, SDM can usually spread AoI packets more evenly than TDM, which could be beneficial as it avoids AoI packet clusters, but it could also lead to a prolonged AoI if the spread update frequency is slightly longer than the incoming update frequency. In that case, TDM, by setting the prioritization periods to directly match the update frequency, could be more useful. However, a key drawback of TDM is that it needs to constantly access system time, which can incur significant overhead. For that, SDM is computationally more favorable than TDM. The advantage of TDM is that it does not need the packet size information. All operations of TDM can be done by only monitoring the link status, which allows one to fully decouple the subqueues from the scheduler. 

With SDM or TDM installed at each outport, the system operator can solve LAC to derive the corresponding AoI ratios $\gamma_l$ for all link $l$, and assign the $\gamma_l$ value to the queueing policy governing the link $l$.

\section{Implementation}\label{sec:implementation}
We demonstrate the feasibility of our proposed IFC scheme by implementing AAQ and \IFIL as Linux kernel modules.
We remark that this is not the only way to potentially implement IFC. For instance, one could instead use servers as classifiers and schedulers. 
We have also successfully implemented IFC in DPDK \cite{dpdk} to demonstrate the potential of a smartNIC deployment. However, DPDK is vendor-specific and its deployment in a virtualized environment (such as cloud-based content-delivery networks) is highly involved. Therefore, we also implement IFC as Linux kernel modules so that it is easy for embedded and commodity Linux systems, such as connected vehicles, networked robotic systems, and public clouds, to import our design without further modification.

\begin{figure}
\centering
\captionbox{Architecture of AoI-aware queueing discipline (AAQ).\label{fig:implementation_architecture_aaq}}{
\includegraphics[scale=0.82]{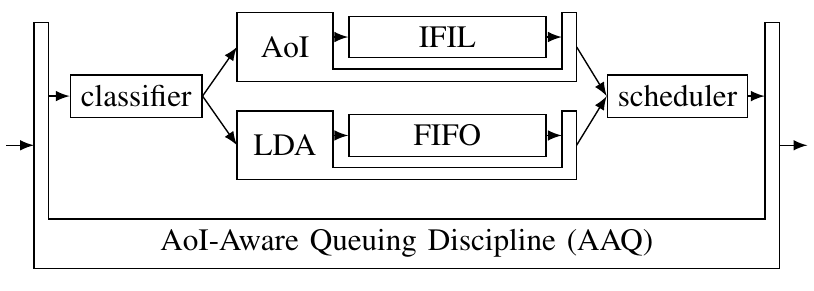}
}\hfill
\captionbox{Architecture of hashed \IFIL. The rectangles are the packets and the boxes marked by m are the meta-data of the packets.\label{fig:implementation_architecture_ifil}}{
\includegraphics[scale=0.82]{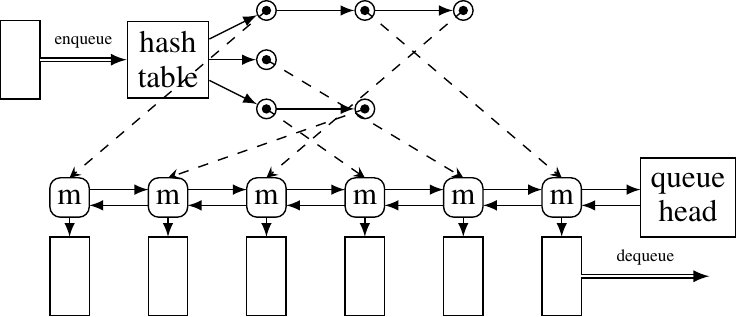}
}
\end{figure}

\paragraph{AAQ}
As shown in \fig{implementation_architecture_aaq}, we pack the classifier and the scheduler into the AAQ module. 
The classifier is a function that categorizes packets based on their header. In our implementation, the classifier distinguishes packets using their layer $4$ protocol number as we deem UDP traffic the AoI traffic in the experiments. One can adopt other classifiers according to their definition of AoI traffic. For instance, one can also design a special layer $3$ protocol for AoI traffic, and the corresponding classifier filters traffic according to the layer $3$ protocol number.

AAQ has two subqueues (or ``classes'' as in Linux traffic control terminology) for LDA and AoI flows. By default, we employ FIFO for the LDA subqueue and \IFIL for the AoI subqueue. We can change the queueing disciplines of the subqueues to handle the flows differently. The scheduler can prioritize these two subqueues according to its scheduling policy, e.g., SDM or TDM. Each time the scheduling policy is invoked, it examines the subqueues and returns one-bit Boolean value to indicate which subqueue the scheduler should dequeue from.

\paragraph{\IFIL}
To enforce \IFIL, outdated packets within the queue need a way of quickly being replaced with fresher arrivals. We could do so by linearly searching through the queue, but this introduces significant overhead.
Instead, we store the packets in the queue as a doubly-linked list of their meta-data and maintain a hash table to locate the packet from each flow. The architecture is shown in \fig{implementation_architecture_ifil}. Since we keep only one packet per flow in the \IFIL queue, we can pre-allocate the memory for the hash items to reduce overhead due to memory allocation. 
We remark that the total number of hash items and meta-data is bounded by the length of the queue. Under moderate queue size, e.g., $10^4$ to $10^5$ packets, the whole data structure (hash items and meta-data) could easily fit in the cache.

\begin{figure}
\centering
\includegraphics{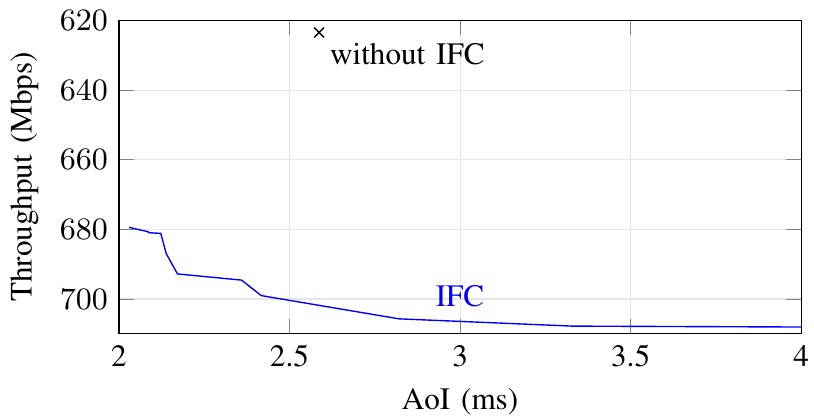}
\caption{The throughput/AoI trade-off of a one-link (physical, $1$ Gbps) example. The results show that our Linux kernel implementation can sustain at least a $1$-Gbps link, providing the IFC benefits without incurring significant overhead.}
\label{fig:implementation_performance}
\end{figure}

\paragraph{Effectiveness}
To demonstrate that our Linux kernel implementation is effective with acceptable overhead, we physically connect two carefully time-synchronized computers by a $1$-Gbps link, pump mixed LDA/AoI traffic through it, and measure the throughput/AoI trade-offs in~\fig{implementation_performance}. The results show that our Linux kernel implementation can serve at least a $1$-Gbps link without incurring significant overhead.
We remark that time-synchronization is only needed because we want to measure the AoI accurately. In practice, we don't need the computers to be time-synchronized to perform IFC.
To gather accurate AoI measurements over multiple computers, we emulate the network and use $100$ Mbps virtual links in \sec{evaluation} to avoid challenging multi-computer time-synchronization.
But given \fig{implementation_performance}, we expect the results extend to physical links of $1$ Gbps or higher bandwidth.

\section{Evaluation}\label{sec:evaluation}
We evaluate FATE and IFC through extensive emulations, reported on in this section. We start with the setup 
in \sec{evaluation-setup} and then address the following questions:
\begin{itemize}
\item \sec{evaluation-overhead}: What is the overhead of performing IFC? How much does our design reduce the overhead?
\item 
\sec{evaluation-rate}: How does FATE perform in comparison to other existing traffic engineering objectives? How does the trade-off factor $\lambda$ in LAC affect the performance?
\item 
\sec{evaluation-queue}: How does IFC perform in comparison to other existing active queue management (AQM)/TCP variants?
\item \sec{evaluation-combination}: Can IFC be combined with existing methods? How much improvement does this provide?
\end{itemize}

\subsection{Setup}\label{sec:evaluation-setup}
We emulate FATE/IFC in Mininet \cite{Mininet}.
Mininet allows us to measure AoI using the same system clock.
We also consider more practical evaluations over real wide-area networks, but synchronization issues make it hard to gather AoI measurements accurately. We remark that the AoI measurements are essential for performance quantification and discussion, but in practice, freshness difference can be experienced directly and so there is no need to measure it.

\paragraph{Network topologies}
We emulate network topologies from Microsoft's SWAN \cite{hong2013achieving}, Internet2 and Google's B4 \cite{jain2013b4, hong2018b4}. 

\paragraph{Existing AQM/TCP Methods}
In addition to IFC with SDM/TDM, we compare against the existing AQM/TCP variants including FIFO (First-In First-Out), LIFO (Last-In First-Out), RED \cite{jacobson1998recommendations}, FQ-CoDel \cite{taht2018flow}, BBR \cite{cardwell2016bbr}, and DCTCP \cite{alizadeh2011data}.
We directly use Linux {\tt tc pfifo}, {\tt red}, {\tt fq\_codel}, {\tt tcp bbr}, and {\tt dctcp} as their implementations using default Linux parameter settings. For pure AQM methods, we pair them with default Linux TCP (CUBIC).

\paragraph{Traffic Characteristics}
In the following experiments, we generate $100$ random traffic patterns with mixed LDA and AoI flows. The LDA and AoI flows are generated by picking each source-destination pair with probability $0.1$ (It is possible for an LDA and an AoI flow to share the same source and destination). We then average the measurements over all generated patterns and plot the average as the final result. 

\paragraph{Evaluation Metrics}
For each generated traffic pattern, we measure the total throughput of the LDA flows and the total AoI of the AoI flows.

\subsection{Overhead of \IFIL and Schedulers}\label{sec:evaluation-overhead}
Since IFC is designed to achieve low AoI, we have to ensure that the overhead is low when adopting it in the system. In particular, we are interested in the overhead of two components: the \IFIL queue and the scheduler.

\begin{figure}
\centering
\includegraphics{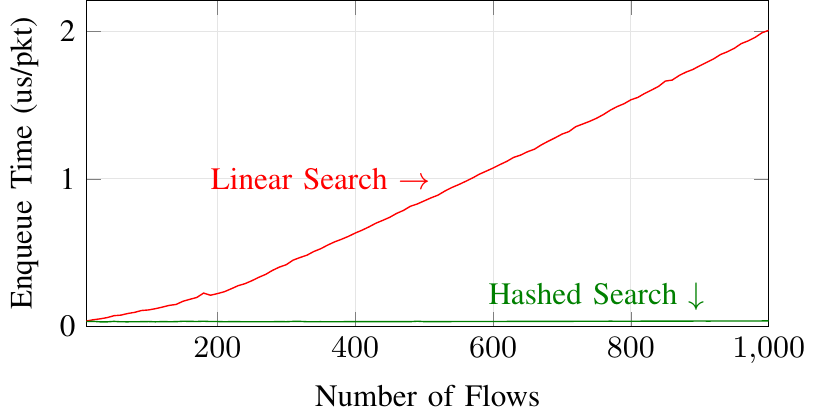}
\vspace*{-0.5\baselineskip}
\caption{The overhead of \IFIL is small with hashed design, about $34.61$ ns per packet. In exchange, we get huge AoI improvement.}
\label{fig:emulation_overhead_lifow}
\end{figure}

We first examine the overhead of the \IFIL queue. When a packet arrives at a \IFIL queue, it replaces the packet of the same flow, if there exists one, which requires searching through the queue. The naive method would be to search linearly through the queue, which can be inefficient when the queue stores a large number of flows.
Instead, we propose hashed implementation in our design. In \fig{emulation_overhead_lifow}, we quantify the overhead by pumping packets from a number of flows into the queue and measuring the enqueue time. \fig{emulation_overhead_lifow} shows that the proposed hashed implementation significantly outperforms linear search. The overhead of the hashed implementation is around $34.61$ ns, while the linear search overhead grows linearly.  Notice that \IFIL keeps only one packet per flow, meaning that we only need to keep a small number (number of flows) of hashed items and packet meta-data to perform hashing, which could fit it the cache.

Another overhead we need to consider is the scheduler overhead. For comparison, we prepare two fully loaded FIFO queues as the AoI and LDA queues and adopt the scheduler to dequeue packets from the two queues.
The scheduler dequeues according to the scheduling policy, and we measure the dequeue overhead in terms of the time.
If we keep dequeueing packets from non-empty queues without scheduling, the overhead is $3.79$ ns (per packet). On the other hand, the dequeue overhead for SDM and TDM are $7.07$ and $21.98$ ns, respectively, regardless of the AoI ratio.
This validates the fact that both SDM and TDM have very low overhead, though TDM requires more time since it inquiries system time per packet arrival.

\subsection{Comparison Against Other Traffic Engineering Objectives}\label{sec:evaluation-rate}
\def\ResultRatea{0.125}
\def\ResultRateb{5}
\def\ResultRatec{49}
\def\ResultRated{71}
\def\ResultRatee{10}
\def\ResultRatef{13}
\def\ResultRateg{7}
\def\ResultRateh{12}
\def\ResultRatei{5}

In addition to FATE that balances sending rate/update frequency through LAC and the trade-off factor $\lambda$, other traffic engineering objectives exist in favor of different trade-offs. In this subsection, we compare FATE with existing traffic engineering objectives and their corresponding rate allocations below. We enforce FATE solution through IFC with SDM or TDM, and the existing schemes run with traditional FIFO queues.

Traditional traffic engineering solves multi-commodity flow (MCF) problem to maximize the overall throughput:
\begin{align*}
\max \sum\limits_{f \in F_\LDA} r_f + \sum\limits_{f \in F_\AoI} \mu_f s_f
\tag{Max Throughput}
\end{align*}
subject to the constraint set of LAC. Without taking AoI into account, Max Throughput achieves good throughput for all flows while not ensuring short AoI for AoI flows. Instead of maximizing the overall throughput, assuming all LDA packets are of the same $s_\LDA$, we can also solve MCF to minimize AoI (proxy) by considering the objective
\begin{align*}
\min \sum\limits_{f \in F_\LDA} \frac{s_\LDA}{2 r_f} + \sum\limits_{f \in F_\AoI} \frac{1}{2 \mu_f}.
\tag{Min AoI}
\end{align*}
Unlike Max Throughput, Min AoI would prefer short AoI for all flows. Some recent papers \cite{lou2020aoi,lou2021boosting} tries to minimize AoI for AoI flows while maximizing their throughput, which lead to the two-objective optimization:
\footnote{\cite{lou2020aoi,lou2021boosting} optimize not only the update frequencies but also the routes for the AoI flows. Accordingly, AoI is given by
\begin{align*}
\frac{1}{2\mu_f} + \sum\limits_{l:l\in f}\frac{s_f}{c_l}.
\end{align*}
Here we predetermine the routes and hence the second term is a constant and omitted from the objective.}
\begin{align}
\min \sum\limits_{f \in F_\AoI} \frac{1}{2\mu_f} \text{\ \ and\ \ }
\max \sum\limits_{f \in F_\AoI} \mu_f s_f.
\tag{Lou 2020}
\end{align}
The algorithm in \cite{lou2020aoi,lou2021boosting} approaches these two objectives iteratively. It first initializes the target total throughput $v$ as $0$. Then, it finds the routing by minimizing AoI with total throughput greater than $v$. Under the resulting routing, it calculates the max achievable throughput and updates $v$ accordingly. The procedure is repeated until a Pareto optimum is reached. In our setting, the routing is fixed and hence the optimum will be obtained after one iteration.

\begin{figure*}
\centering
\includegraphics[scale=0.8]{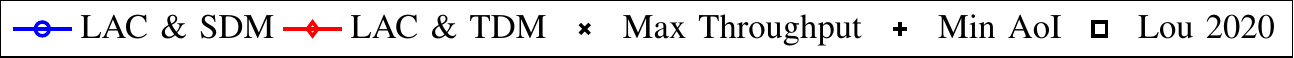}\\
\subcaptionbox{Network topology: B4}[\textwidth][b]{
\includegraphics[scale=0.8]{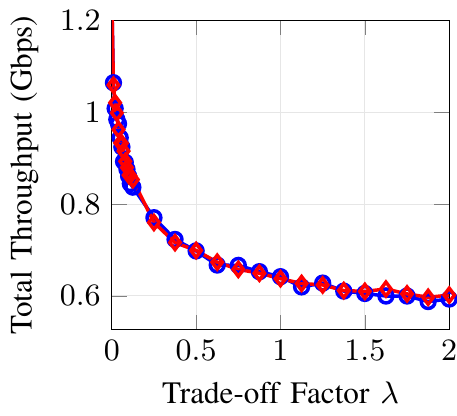}\hfill
\includegraphics[scale=0.8]{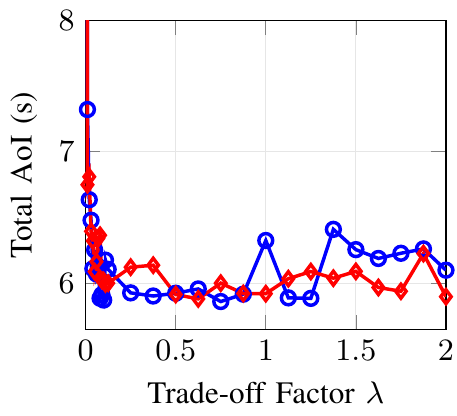}\hfill
\includegraphics[scale=0.8]{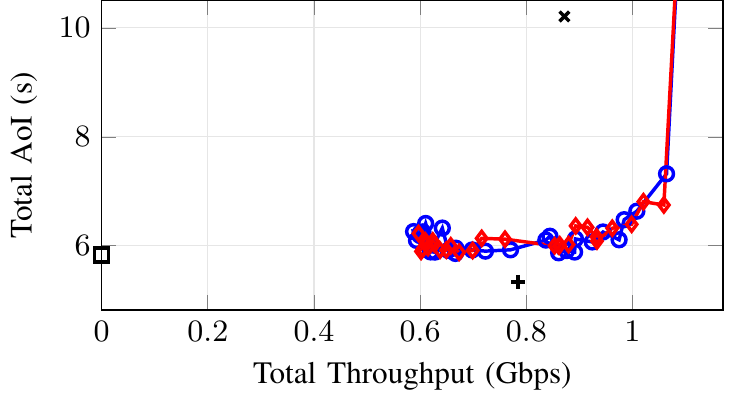}
}\\
\subcaptionbox{Network topology: SWAN}[\textwidth][b]{
\includegraphics[scale=0.8]{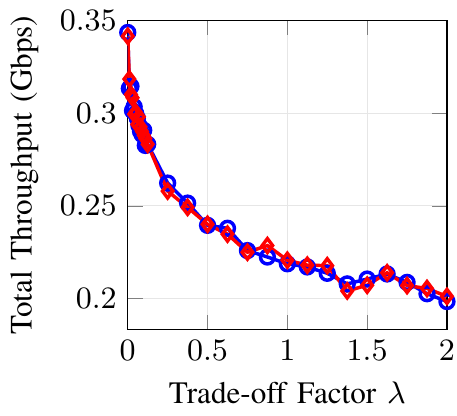}\hfill
\includegraphics[scale=0.8]{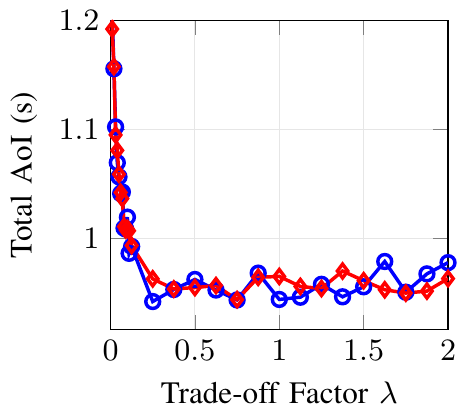}\hfill
\includegraphics[scale=0.8]{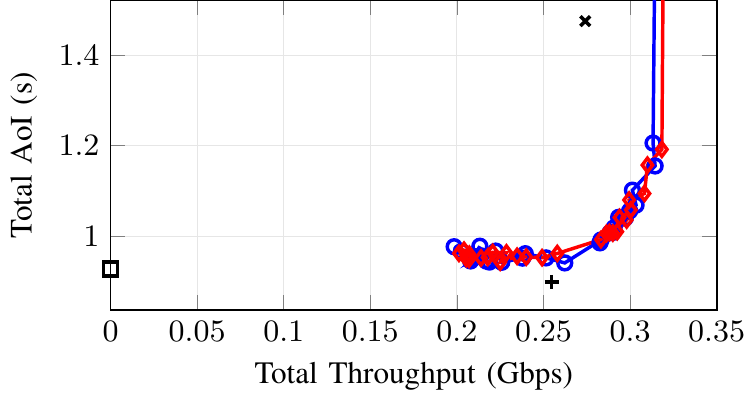}
}
\caption{We compare the rate allocation by FATE with other rate allocations from different traffic engineering objectives (Max Throughput, Min AoI, and Lou 2020) under two different network topologies. FATE solves LAC to trade throughput for AoI according to $\lambda$, and we plot the decreasing of throughput/AoI against $\lambda$ on the left. Enforcing LAC solution via either SDM or TDM as the IFC gives similar performance. On the right we examine the AoI-throughput trade-off. LAC can achieve shorter AoI than traditional thorughput-only scheme, while achieving higher throughput than AoI emphasized schemes.
}
\label{fig:comparison-LAC}
\end{figure*}

In \fig{comparison-LAC}, we examine the how the trade-off factor $\lambda$ influences total throughput and total AoI on the left two plots and what trade-offs different methods achieve on the right under the topologies B4 and SWAN. 
As expected, we can see that total throughput and total AoI both decrease when the trade-off factor $\lambda$ increases.
Comparing against other methods (under $\lambda = \ResultRatea$), Max Throughput achieves up to $\ResultRateb \%$ more throughput while the AoI is $\ResultRatec$ to $\ResultRated \%$ longer than LAC.
Min AoI achieves the best AoI, which is $\ResultRatee$ to $\ResultRatef \%$ shorter than LAC, while the maintaining a relatively good throughput. It is because a low AoI would also lead to high throughput for the LDA flows. Even through, LAC can choose to trade slightly more AoI for even higher throughput -- $\ResultRateg$ to $\ResultRateh \%$ higher than Min AoI -- by setting smaller $\lambda$. On the other hand, Lou 2020 also achieves small AoI (about $\ResultRatei \%$ shorter than LAC), but it focuses only on AoI flows' AoI and hence LDA flows are allocated zero sending rates.

\subsection{Comparison Against Other AQM/TCP Variants}\label{sec:evaluation-queue}
\def\ResultQueuea{0.125}
\def\ResultQueueb{33}
\def\ResultQueuec{38}
\def\ResultQueued{47}
\def\ResultQueuee{60}
\def\ResultQueuef{33}
\def\ResultQueueg{19}
\def\ResultQueueh{30}
\def\ResultQueuei{21}
\def\ResultQueuej{26}

\begin{figure}
\centering
\includegraphics[scale=0.8]{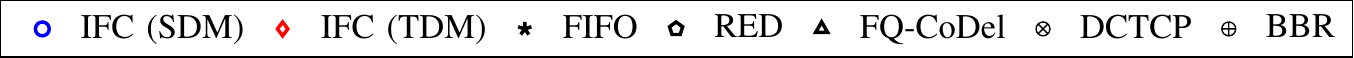}\\[0.5\baselineskip]
\subcaptionbox{Network topology: B4}{
\includegraphics[scale=0.8]{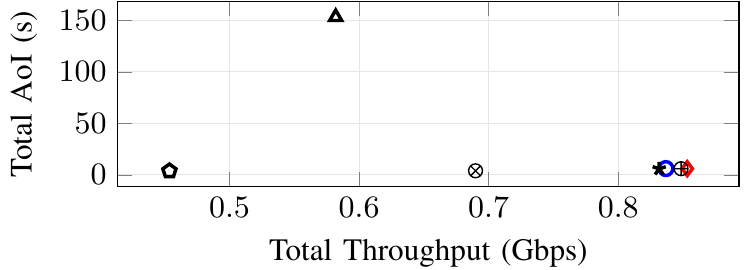}
}\hfill
\subcaptionbox{Network topology: SWAN}{
\includegraphics[scale=0.8]{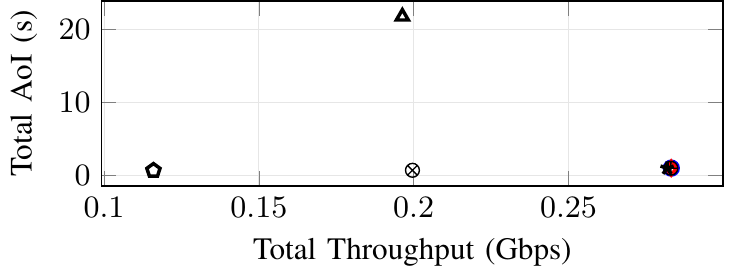}
}
\caption{Under the same rate/update frequency, IFC can achieve shorter AoI and higher throughput than some state-of-the-art queueing management methods by scheduling and enforcing different queueing policies to LDA and AoI flows at each node in the network.}
\label{fig:comparison-queues}
\end{figure}

After we assign the sending rate/update frequency according to the FATE solution, IFC takes over and performs packet scheduling at the nodes. In the following, we assign sending rate/update frequencies derived from LAC (under $\lambda = \ResultQueuea$) and compare the performance achieved by IFC against other existing AQM/TCP variants, including FIFO, RED, FQ-CoDel, BBR, and DCTCP. The results are shown in \fig{comparison-queues}.

\fig{comparison-queues} shows that IFC, FIFO, and BBR achieve the highest throughput with the lowest AoI. RED drops packets more aggressively to maintain a short queue, which leads to a very short AoI ($\ResultQueueb$ to $\ResultQueuec \%$ shorter than IFC). On the other hand, aggressive packet drop significantly hurts the throughput, which degrades to $\ResultQueued$ to $\ResultQueuee\%$ lower. In comparison, DCTCP and BBR maintain short queues through sending rate control and hence the throughput is higher than RED while keeping AoI low. Between the two TCP protocols, BBR saturates the existing bandwidth better than DCTCP. DCTCP achieves about $\ResultQueuef \%$ shorter AoI with $\ResultQueueg$ to $\ResultQueueh \%$ lower throughput.
FQ-CoDel tries to allocate bandwidth to LDA and AoI flows more evenly, which obtains slightly higher throughput than RED but suffers long AoI significantly, which is $\ResultQueuei$ to $\ResultQueuej \times$ longer than IFC.

\subsection{Combination with Other Methods}\label{sec:evaluation-combination}
\def\ResultCombinationa{42}
\def\ResultCombinationb{44}
\def\ResultCombinationc{22}
\def\ResultCombinationd{25}
\def\ResultCombinatione{50}
\def\ResultCombinationf{1.9}
\def\ResultCombinationg{2.4}
\def\ResultCombinationh{42}
\def\ResultCombinationi{46}
\def\ResultCombinationj{23}
\def\ResultCombinationk{40}

\begin{figure*}
\centering
\includegraphics[scale=0.8]{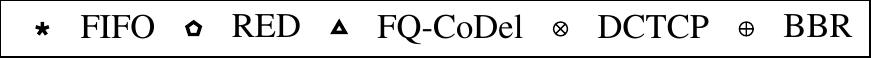}\\
\subcaptionbox{Network topology: B4}{
\includegraphics[scale=0.8]{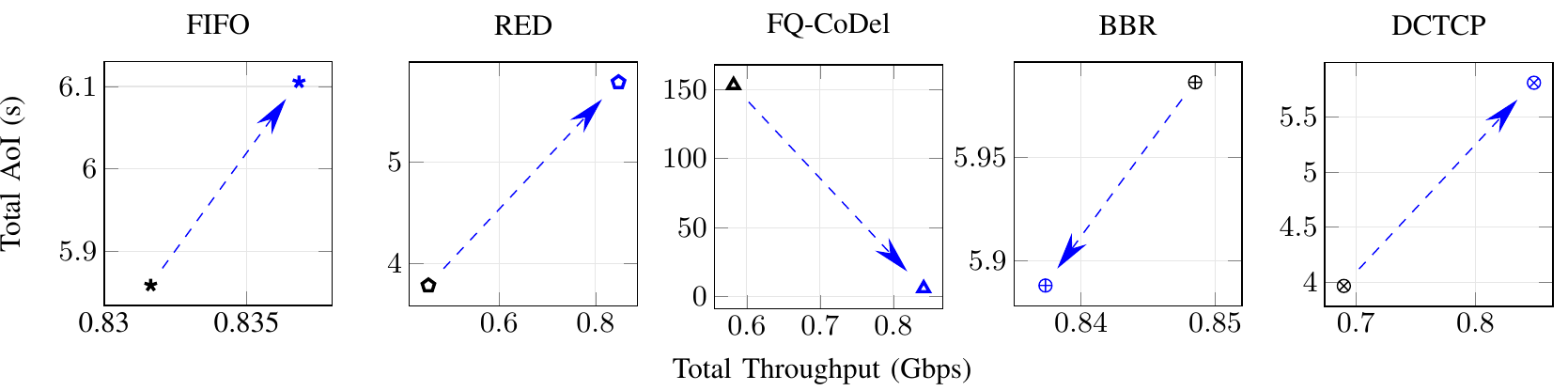}
}\\
\subcaptionbox{Network topology: SWAN}{
\includegraphics[scale=0.8]{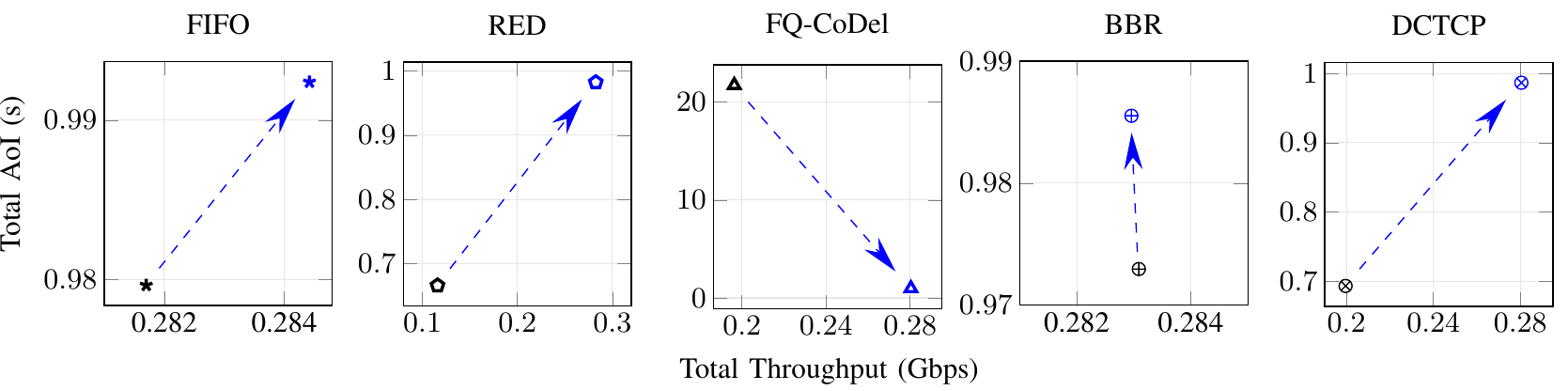}
}
\caption{In the default design of IFC, we use FIFO for the LDA flows. 
We can use different queueing discipline for LDA flows. Under the same rate allocation, we compare the performance of some queueing disciplines (black) with their combination with IFC (blue). After combining with IFC, we have better performance (towards the bottom/right) or achieve different AoI-throughput trade-offs (towards the top right) under both networks.
}
\label{fig:comparison-combination}
\end{figure*}

A key benefit of IFC is that it can easily be combined with other methods to improve their performance. Specifically, we can replace the FIFO queue in IFC with a different queueing discipline in order to handle LDA flows differently. In \fig{comparison-combination}, we examine the performance improvements of FIFO, RED, FQ-CoDel, DCTCP, BBR and  after combining them with IFC (with scheduler SDM) under different network topologies.

We draw an arrow to show the performance of the method before and after combined with IFC. It is better if the arrow goes down or to the right. 
After combination, IFC achieves better performance than fair queueing based FQ-CoDel ($\ResultCombinationa$ to $\ResultCombinationb\%$ more throughput with $\ResultCombinationc$ to $\ResultCombinationd \times$ shorter AoI). It trades AoI for throughput when combining with low queue length methods RED (about $\ResultCombinatione\%$ longer AoI for $\ResultCombinationf$ to $\ResultCombinationg \times$ larger throughput) and DCTCP ($\ResultCombinationh$ to $\ResultCombinationi \%$ longer AoI for $\ResultCombinationj$ to $\ResultCombinationk \%$ more throughput). After combining with FIFO and BBR, IFC performs well when the network is much more congested as in B4. In SWAN, the network is less congested and hence FIFO and BBR performs similar to IFC.

\section{Related Work}\label{sec:related-work}
Data freshness is a key requirement to achieve safety and stability in many communities, including connected vehicles/satellites \cite{gandhi2007pedestrian, bandyop16, talak2016speed, biomo2014routing, hoel2019combining, talak2019optimizing}, real-time database \cite{adelberg1995applying,kang2004managing,kim2016sporadic}, robotics \cite{lawton00,bandyop17probswarm,tardioli16,subTwebsite}, and power systems \cite{phadke1993synchronized,naduvathuparambil2002communication,terzija2011wide,ghosh2013communication,celli2014dms,primadianto2016review}. 
State-of-the-art designs use networks for connectivity only and strive for freshness at the end systems by active probing \cite{hu2015data,kim2016sporadic,dhekne2019trackio}, dedicated links \cite{naduvathuparambil2002communication,ghosh2013communication}, or local decision \cite{hoel2019combining}.

The goal of improving freshness of flows in a network has also captured the interest of the information theory community over the past decade since the initial work of \cite{kaul2011minimizing,kaul2012real} introduced the concept of the AoI. 
So far, most research papers in this area focus on theoretical analysis and consider the updates generated by a stochastic source going through different queueing systems, including one-queue systems ($M/M/1$, $M/D/1$, $D/M/1$ \cite{kaul2012real}, $M/M/1/1$, $M/M/1/2*$ \cite{costa2014age,costa2016age}, $M/G/1$, $M/G/1/1$ \cite{huang2015optimizing}, and multiple sources \cite{yates2012real}), multihop networks \cite{bedewy2017age, talak2018scheduling, talak2019optimizing, bedewy2019minimizing,lou2020aoi,lou2021boosting}, and stochastic network models \cite{kam2013age,kavitha2018controlling,lu2018age}.
Some recent efforts \cite{franco2016lupmac,shreedhar2018acp,kadota2020age} start investigating practical protocols for fresh AoI.
This literature has provided many insights for design. For example, when multiple sources are present, \cite{pappas2015age} proposes to keep only the freshest update in the queue and simulation results demonstrate that the approach can effectively reduce the queueing delay. Another example is that the zero-wait policy, a.k.a., the work-conservation policy, is usually not AoI optimal \cite{yates2015lazy,sun2016update,sun2017update}. Instead, a ``lazy'' policy performs better \cite{yates2015lazy}.
We refer the reader to \cite{yates2019age, lou2021boosting} for more comprehensive surveys on the prior work on AoI.

One recent set of papers that is related to ours focuses on analytic studies of the interplay between freshness and traditional measures, like throughput and energy. For example,\cite{bedewy2016optimizing,kadota2018optimizing,kadota2019scheduling,
lou2020aoi,lou2021boosting} focus on scheduling and queueing disciplines for flows that care about a combination of freshness and throughput. 
\cite{altman2019forever,ceran2019average} study the energy cost of maintaining freshness.
We remark that the problems investigated in these papers are different from ours in that they impose constraints to maintain throughput/energy level for the AoI flows. In contrast, in our work, the AoI flows are interested in AoI \emph{only}, and a separate class of flows (LDA flows) is interested in throughput. 

Our two-layer approach to freshness can be attributed to two main ideas: 
active queue management and heterogeneous flow scheduling. Active queue management methods control the queue size through packet drop or explicit congestion notification. They aim to avoid congestion \cite{jacobson1998recommendations,feng1999self} and bufferbloat \cite{nichols2012controlling,pan2013pie,taht2018flow}. Although those methods also drop packets, they do not consider the heterogeneity of flows, neither do they differentiate packets from the same flow according to their properties such as freshness. 

On the other hand, heterogeneous flow scheduling systems acknowledge the variety of objectives among flows and exploit their differences to get better performance. For instance, HULL \cite{alizadeh2012less} deals with throughput and ultra-low latency applications. pFabric \cite{alizadeh2013pfabric} and PIAS \cite{bai2015information} reduce flow completion time by differentiating short and long flows. D$^3$ \cite{wilson2011better} satisfies deadlines of alive flows by quenching the flows past due. NetStitcher \cite{laoutaris2011inter} improves average utilization by manipulating bulk traffic. All of the systems above leverage the diversity of the flows to achieve their design goals, and our design joins this line of research as the first to address the trade-off between throughput and freshness.

\section{Conclusion}\label{sec:conclusion}
We investigate the trade-offs between LDA flows' throughput and AoI flows' AoI in wired networks. Our approach to the problem consists of two layers, the freshness-aware traffic engineering (FATE) at the higher layer and the in-network freshness control (IFC) at the lower layer. We formulate the LDA-AoI Coscheduling (LAC) problem to derive the desired sending rate/update frequency to maximize the throughput of LDA flows and an approximated AoI of the AoI flows. We justify our approximation by showing the NP-hardness of zero-queueing schedule and providing an upper bound on AoI. The LAC solution is then enforced by IFC through AoI-Aware Queueing (AAQ). We propose two scheduling policies for AAQ -- the size-driven multiplexing (SDM) and time-division multiplexing (TDM), and we provide efficient Linux IFC implementations. Through extensive emulations, we show that our IFC implementation has low overhead and we can potentially trade a little throughput ($\ResultRateb\%$) for much shorter AoI ($\ResultRatec$ to $\ResultRated\%$ shorter) comparing to traditional throughput-based objective. Also, we can combine IFC with existing AQM/TCP variants to achieve better performance or different AoI-throughput trade-offs.

\label{theEndofContent}

\newcommand{\shownote}[1]{\unskip}
\newcommand{\showURL}[1]{#1.}
\bibliographystyle{ACM-Reference-Format-num}
\bibliography{Test}


\begin{thebibliography}{74}


\ifx \showCODEN    \undefined \def \showCODEN     #1{\unskip}     \fi
\ifx \showDOI      \undefined \def \showDOI       #1{#1}\fi
\ifx \showISBNx    \undefined \def \showISBNx     #1{\unskip}     \fi
\ifx \showISBNxiii \undefined \def \showISBNxiii  #1{\unskip}     \fi
\ifx \showISSN     \undefined \def \showISSN      #1{\unskip}     \fi
\ifx \showLCCN     \undefined \def \showLCCN      #1{\unskip}     \fi
\ifx \shownote     \undefined \def \shownote      #1{#1}          \fi
\ifx \showarticletitle \undefined \def \showarticletitle #1{#1}   \fi
\ifx \showURL      \undefined \def \showURL       {\relax}        \fi
\providecommand\bibfield[2]{#2}
\providecommand\bibinfo[2]{#2}
\providecommand\natexlab[1]{#1}
\providecommand\showeprint[2][]{arXiv:#2}

\bibitem[\protect\citeauthoryear{??}{sub}{[n. d.]}]%
        {subTwebsite}
\bibinfo{title}{{DARPA} {S}ubterranean ({SubT}) Challenge}.
\newblock
\showURL{%
\url{https://www.subtchallenge.com/}}
\newblock
\shownote{Accessed: 2020-02-20.}


\bibitem[\protect\citeauthoryear{??}{dpd}{[n. d.]}]%
        {dpdk}
\bibinfo{title}{Data Plane Development Kit}.
\newblock
\showURL{%
\url{http://dpdk.org}}


\bibitem[\protect\citeauthoryear{??}{Fac}{[n. d.]}]%
        {Facebook-Live}
\bibinfo{title}{Facebook Live}.
\newblock
\showURL{%
\url{https://live.fb.com}}


\bibitem[\protect\citeauthoryear{??}{Goo}{[n. d.]}]%
        {Google-Stadia-Cloud-Gaming}
\bibinfo{title}{Google {Stadia}}.
\newblock
\showURL{%
\url{https://store.google.com/magazine/stadia}}
\newblock
\shownote{Cloud gaming platform.}


\bibitem[\protect\citeauthoryear{??}{Min}{[n. d.]}]%
        {Mininet}
\bibinfo{title}{Mininet}.
\newblock
\showURL{%
\url{http://mininet.org/}}


\bibitem[\protect\citeauthoryear{??}{You}{[n. d.]}]%
        {YouTube-TV}
\bibinfo{title}{{YouTube} {TV}}.
\newblock
\showURL{%
\url{https://tv.youtube.com}}


\bibitem[\protect\citeauthoryear{Adelberg, Garcia-Molina, and Kao}{Adelberg
  et~al\mbox{.}}{1995}]%
        {adelberg1995applying}
\bibfield{author}{\bibinfo{person}{Brad Adelberg}, \bibinfo{person}{Hector
  Garcia-Molina}, {and} \bibinfo{person}{Ben Kao}.}
  \bibinfo{year}{1995}\natexlab{}.
\newblock \showarticletitle{Applying Update Streams in a Soft Real-Time
  Database System}. In \bibinfo{booktitle}{{\em Proc. {ACM} {SIGMOD}}},
  Vol.~\bibinfo{volume}{24}. ACM, \bibinfo{pages}{245--256}.
\newblock


\bibitem[\protect\citeauthoryear{Alizadeh, Greenberg, Maltz, Padhye, Patel,
  et~al\mbox{.}}{Alizadeh et~al\mbox{.}}{2011}]%
        {alizadeh2011data}
\bibfield{author}{\bibinfo{person}{Mohammad Alizadeh}, \bibinfo{person}{Albert
  Greenberg}, \bibinfo{person}{David~A Maltz}, \bibinfo{person}{Jitendra
  Padhye}, \bibinfo{person}{Parveen Patel}, {et~al\mbox{.}}}
  \bibinfo{year}{2011}\natexlab{}.
\newblock \showarticletitle{Data Center {TCP} ({DCTCP})}.
\newblock \bibinfo{journal}{{\em {ACM} {SIGCOMM} {CCR}\/}}
  \bibinfo{volume}{41}, \bibinfo{number}{4} (\bibinfo{year}{2011}),
  \bibinfo{pages}{63--74}.
\newblock


\bibitem[\protect\citeauthoryear{Alizadeh, Kabbani, Edsall, Prabhakar, Vahdat,
  et~al\mbox{.}}{Alizadeh et~al\mbox{.}}{2012}]%
        {alizadeh2012less}
\bibfield{author}{\bibinfo{person}{Mohammad Alizadeh}, \bibinfo{person}{Abdul
  Kabbani}, \bibinfo{person}{Tom Edsall}, \bibinfo{person}{Balaji Prabhakar},
  \bibinfo{person}{Amin Vahdat}, {et~al\mbox{.}}}
  \bibinfo{year}{2012}\natexlab{}.
\newblock \showarticletitle{Less is More: Trading a little Bandwidth for
  Ultra-Low Latency in the Data Center}. In \bibinfo{booktitle}{{\em Proc.
  {USENIX} {NSDI}}}.
\newblock


\bibitem[\protect\citeauthoryear{Alizadeh, Yang, Sharif, Katti, McKeown,
  et~al\mbox{.}}{Alizadeh et~al\mbox{.}}{2013}]%
        {alizadeh2013pfabric}
\bibfield{author}{\bibinfo{person}{Mohammad Alizadeh}, \bibinfo{person}{Shuang
  Yang}, \bibinfo{person}{Milad Sharif}, \bibinfo{person}{Sachin Katti},
  \bibinfo{person}{Nick McKeown}, {et~al\mbox{.}}}
  \bibinfo{year}{2013}\natexlab{}.
\newblock \showarticletitle{{pFabric}: Minimal Near-Optimal Datacenter
  Transport}.
\newblock \bibinfo{journal}{{\em {ACM} {SIGCOMM} {CCR}\/}}
  \bibinfo{volume}{43}, \bibinfo{number}{4} (\bibinfo{year}{2013}),
  \bibinfo{pages}{435--446}.
\newblock


\bibitem[\protect\citeauthoryear{Altman, El-Azouzi, Menasche, and Xu}{Altman
  et~al\mbox{.}}{2019}]%
        {altman2019forever}
\bibfield{author}{\bibinfo{person}{Eitan Altman}, \bibinfo{person}{Rachid
  El-Azouzi}, \bibinfo{person}{Daniel~Sadoc Menasche}, {and}
  \bibinfo{person}{Yuedong Xu}.} \bibinfo{year}{2019}\natexlab{}.
\newblock \showarticletitle{Forever Young: Aging Control for Hybrid Networks}.
  In \bibinfo{booktitle}{{\em Proc. {ACM} {Mobihoc}}}.
  \bibinfo{pages}{91--100}.
\newblock


\bibitem[\protect\citeauthoryear{Bai, Chen, Wang, Chen, Han, et~al\mbox{.}}{Bai
  et~al\mbox{.}}{2015}]%
        {bai2015information}
\bibfield{author}{\bibinfo{person}{Wei Bai}, \bibinfo{person}{Kai Chen},
  \bibinfo{person}{Hao Wang}, \bibinfo{person}{Li Chen},
  \bibinfo{person}{Dongsu Han}, {et~al\mbox{.}}}
  \bibinfo{year}{2015}\natexlab{}.
\newblock \showarticletitle{Information-Agnostic Flow Scheduling for Commodity
  Data Centers}. In \bibinfo{booktitle}{{\em Proc. {USENIX} {NSDI}}}.
  \bibinfo{pages}{455--468}.
\newblock


\bibitem[\protect\citeauthoryear{{Bandyopadhyay}, {Chung}, and
  {Hadaegh}}{{Bandyopadhyay} et~al\mbox{.}}{2017}]%
        {bandyop17probswarm}
\bibfield{author}{\bibinfo{person}{S. {Bandyopadhyay}}, \bibinfo{person}{S.
  {Chung}}, {and} \bibinfo{person}{F.~Y. {Hadaegh}}.}
  \bibinfo{year}{2017}\natexlab{}.
\newblock \showarticletitle{Probabilistic and Distributed Control of a
  Large-Scale Swarm of Autonomous Agents}.
\newblock \bibinfo{journal}{{\em IEEE Trans. Robot.\/}} \bibinfo{volume}{33},
  \bibinfo{number}{5} (\bibinfo{date}{Oct} \bibinfo{year}{2017}),
  \bibinfo{pages}{1103--1123}.
\newblock


\bibitem[\protect\citeauthoryear{Bandyopadhyay, Foust, Subramanian, Chung, and
  Hadaegh}{Bandyopadhyay et~al\mbox{.}}{2016}]%
        {bandyop16}
\bibfield{author}{\bibinfo{person}{Saptarshi Bandyopadhyay},
  \bibinfo{person}{Rebecca Foust}, \bibinfo{person}{Giri Subramanian},
  \bibinfo{person}{Soon-Jo Chung}, {and} \bibinfo{person}{Fred Hadaegh}.}
  \bibinfo{year}{2016}\natexlab{}.
\newblock \showarticletitle{Review of Formation Flying and Constellation
  Missions Using Nanosatellites}.
\newblock \bibinfo{journal}{{\em J. Spacecraft Rockets\/}}
  \bibinfo{volume}{53} (\bibinfo{date}{Mar} \bibinfo{year}{2016}),
  \bibinfo{pages}{1--12}.
\newblock


\bibitem[\protect\citeauthoryear{Bedewy, Sun, and Shroff}{Bedewy
  et~al\mbox{.}}{2016}]%
        {bedewy2016optimizing}
\bibfield{author}{\bibinfo{person}{Ahmed~M Bedewy}, \bibinfo{person}{Yin Sun},
  {and} \bibinfo{person}{Ness~B Shroff}.} \bibinfo{year}{2016}\natexlab{}.
\newblock \showarticletitle{Optimizing Data Freshness, Throughput, and Delay in
  Multi-Server Information-Update Systems}. In \bibinfo{booktitle}{{\em Proc.
  {IEEE} {ISIT}}}. IEEE, \bibinfo{pages}{2569--2573}.
\newblock


\bibitem[\protect\citeauthoryear{Bedewy, Sun, and Shroff}{Bedewy
  et~al\mbox{.}}{2017}]%
        {bedewy2017age}
\bibfield{author}{\bibinfo{person}{Ahmed~M Bedewy}, \bibinfo{person}{Yin Sun},
  {and} \bibinfo{person}{Ness~B Shroff}.} \bibinfo{year}{2017}\natexlab{}.
\newblock \showarticletitle{Age-Optimal Information Updates in Multihop
  Networks}. In \bibinfo{booktitle}{{\em Proc. {IEEE} {ISIT}}}. IEEE,
  \bibinfo{pages}{576--580}.
\newblock


\bibitem[\protect\citeauthoryear{Bedewy, Sun, and Shroff}{Bedewy
  et~al\mbox{.}}{2019}]%
        {bedewy2019minimizing}
\bibfield{author}{\bibinfo{person}{Ahmed~M Bedewy}, \bibinfo{person}{Yin Sun},
  {and} \bibinfo{person}{Ness~B Shroff}.} \bibinfo{year}{2019}\natexlab{}.
\newblock \showarticletitle{Minimizing the Age of Information Through Queues}.
\newblock \bibinfo{journal}{{\em {IEEE} Trans. Inf. Theory\/}}
  \bibinfo{volume}{65}, \bibinfo{number}{8} (\bibinfo{year}{2019}),
  \bibinfo{pages}{5215--5232}.
\newblock


\bibitem[\protect\citeauthoryear{Biomo, Kunz, and St-Hilaire}{Biomo
  et~al\mbox{.}}{2014}]%
        {biomo2014routing}
\bibfield{author}{\bibinfo{person}{Jean-Daniel Medjo~Me Biomo},
  \bibinfo{person}{Thomas Kunz}, {and} \bibinfo{person}{Marc St-Hilaire}.}
  \bibinfo{year}{2014}\natexlab{}.
\newblock \showarticletitle{Routing in Unmanned Aerial Ad Hoc Networks:
  Introducing a Route Reliability Criterion}. In \bibinfo{booktitle}{{\em Proc.
  IFIP WMNC}}. IEEE, \bibinfo{pages}{1--7}.
\newblock


\bibitem[\protect\citeauthoryear{Cardwell, Cheng, Gunn, Yeganeh, and
  Jacobson}{Cardwell et~al\mbox{.}}{2016}]%
        {cardwell2016bbr}
\bibfield{author}{\bibinfo{person}{Neal Cardwell}, \bibinfo{person}{Yuchung
  Cheng}, \bibinfo{person}{C~Stephen Gunn}, \bibinfo{person}{Soheil~Hassas
  Yeganeh}, {and} \bibinfo{person}{Van Jacobson}.}
  \bibinfo{year}{2016}\natexlab{}.
\newblock \showarticletitle{{BBR}: Congestion-based Congestion Control}.
\newblock \bibinfo{journal}{{\em Queue\/}} \bibinfo{volume}{14},
  \bibinfo{number}{5} (\bibinfo{year}{2016}), \bibinfo{pages}{50}.
\newblock


\bibitem[\protect\citeauthoryear{Celli, Pegoraro, Pilo, Pisano, and
  Sulis}{Celli et~al\mbox{.}}{2014}]%
        {celli2014dms}
\bibfield{author}{\bibinfo{person}{Gianni Celli},
  \bibinfo{person}{Paolo~Atillio Pegoraro}, \bibinfo{person}{Fabrizio Pilo},
  \bibinfo{person}{Giuditta Pisano}, {and} \bibinfo{person}{Sara Sulis}.}
  \bibinfo{year}{2014}\natexlab{}.
\newblock \showarticletitle{{DMS} Cyber-Physical Simulation for Assessing the
  Impact of State Estimation and Communication Media in Smart Grid Operation}.
\newblock \bibinfo{journal}{{\em IEEE Trans. Power Syst.\/}}
  \bibinfo{volume}{29}, \bibinfo{number}{5} (\bibinfo{year}{2014}),
  \bibinfo{pages}{2436--2446}.
\newblock


\bibitem[\protect\citeauthoryear{Ceran, G{\"u}nd{\"u}z, and Gy{\"o}rgy}{Ceran
  et~al\mbox{.}}{2019}]%
        {ceran2019average}
\bibfield{author}{\bibinfo{person}{Elif~Tu{\u{g}}{\c{c}}e Ceran},
  \bibinfo{person}{Deniz G{\"u}nd{\"u}z}, {and} \bibinfo{person}{Andr{\'a}s
  Gy{\"o}rgy}.} \bibinfo{year}{2019}\natexlab{}.
\newblock \showarticletitle{Average Age of Information with Hybrid {ARQ} Under
  a Resource Constraint}.
\newblock \bibinfo{journal}{{\em IEEE Trans. Wireless Commun.\/}}
  \bibinfo{volume}{18}, \bibinfo{number}{3} (\bibinfo{year}{2019}),
  \bibinfo{pages}{1900--1913}.
\newblock


\bibitem[\protect\citeauthoryear{{Cisco Visual Networking Index}}{{Cisco Visual
  Networking Index}}{2019}]%
        {cisco2019global}
\bibfield{author}{\bibinfo{person}{{Cisco Visual Networking Index}}.}
  \bibinfo{year}{2019}\natexlab{}.
\newblock \showarticletitle{Global Mobile Data Traffic Forecast Update, 2017 --
  2022}.
\newblock \bibinfo{journal}{{\em Cisco White Paper\/}} (\bibinfo{year}{2019}).
\newblock


\bibitem[\protect\citeauthoryear{Costa, Codreanu, and Ephremides}{Costa
  et~al\mbox{.}}{2014}]%
        {costa2014age}
\bibfield{author}{\bibinfo{person}{Maice Costa}, \bibinfo{person}{Marian
  Codreanu}, {and} \bibinfo{person}{Anthony Ephremides}.}
  \bibinfo{year}{2014}\natexlab{}.
\newblock \showarticletitle{Age of Information with Packet Management}. In
  \bibinfo{booktitle}{{\em Proc. {IEEE} {ISIT}}}. IEEE,
  \bibinfo{pages}{1583--1587}.
\newblock


\bibitem[\protect\citeauthoryear{Costa, Codreanu, and Ephremides}{Costa
  et~al\mbox{.}}{2016}]%
        {costa2016age}
\bibfield{author}{\bibinfo{person}{Maice Costa}, \bibinfo{person}{Marian
  Codreanu}, {and} \bibinfo{person}{Anthony Ephremides}.}
  \bibinfo{year}{2016}\natexlab{}.
\newblock \showarticletitle{On the Age of Information in Status Update Systems
  with Packet Management}.
\newblock \bibinfo{journal}{{\em {IEEE} Trans. Inf. Theory\/}}
  \bibinfo{volume}{62}, \bibinfo{number}{4} (\bibinfo{year}{2016}),
  \bibinfo{pages}{1897--1910}.
\newblock


\bibitem[\protect\citeauthoryear{Dhekne, Chakraborty, Sundaresan, and
  Rangarajan}{Dhekne et~al\mbox{.}}{2019}]%
        {dhekne2019trackio}
\bibfield{author}{\bibinfo{person}{Ashutosh Dhekne}, \bibinfo{person}{Ayon
  Chakraborty}, \bibinfo{person}{Karthikeyan Sundaresan}, {and}
  \bibinfo{person}{Sampath Rangarajan}.} \bibinfo{year}{2019}\natexlab{}.
\newblock \showarticletitle{{TrackIO}: Tracking First Responders Inside-Out}.
  In \bibinfo{booktitle}{{\em Proc. {USENIX} {NSDI}}}.
  \bibinfo{pages}{751--764}.
\newblock


\bibitem[\protect\citeauthoryear{Feng, Kandlur, Saha, and Shin}{Feng
  et~al\mbox{.}}{1999}]%
        {feng1999self}
\bibfield{author}{\bibinfo{person}{Wu-Chang Feng}, \bibinfo{person}{Dilip~D
  Kandlur}, \bibinfo{person}{Debanjan Saha}, {and} \bibinfo{person}{Kang~G
  Shin}.} \bibinfo{year}{1999}\natexlab{}.
\newblock \showarticletitle{A Self-Configuring {RED} Gateway}. In
  \bibinfo{booktitle}{{\em Proc. {IEEE} {INFOCOM}}}. IEEE,
  \bibinfo{pages}{1320--1328}.
\newblock


\bibitem[\protect\citeauthoryear{Franco, Fitzgerald, Landfeldt, Pappas, and
  Angelakis}{Franco et~al\mbox{.}}{2016}]%
        {franco2016lupmac}
\bibfield{author}{\bibinfo{person}{Antonio Franco}, \bibinfo{person}{Emma
  Fitzgerald}, \bibinfo{person}{Bj{\"o}rn Landfeldt}, \bibinfo{person}{Nikolaos
  Pappas}, {and} \bibinfo{person}{Vangelis Angelakis}.}
  \bibinfo{year}{2016}\natexlab{}.
\newblock \showarticletitle{{LUPMAC}: A Cross-Layer {MAC} Technique to Improve
  the Age of Information Over Dense {WLANs}}. In \bibinfo{booktitle}{{\em Proc.
  {IEEE} {ICT}}}. \bibinfo{pages}{1--6}.
\newblock


\bibitem[\protect\citeauthoryear{Gandhi and Trivedi}{Gandhi and
  Trivedi}{2007}]%
        {gandhi2007pedestrian}
\bibfield{author}{\bibinfo{person}{Tarak Gandhi} {and}
  \bibinfo{person}{Mohan~Manubhai Trivedi}.} \bibinfo{year}{2007}\natexlab{}.
\newblock \showarticletitle{Pedestrian Protection Systems: Issues, Survey, and
  Challenges}.
\newblock \bibinfo{journal}{{\em IEEE Trans. Intell. Transp. Syst.\/}}
  \bibinfo{volume}{8}, \bibinfo{number}{3} (\bibinfo{year}{2007}),
  \bibinfo{pages}{413--430}.
\newblock


\bibitem[\protect\citeauthoryear{Ghosh, Ghose, and Mohanta}{Ghosh
  et~al\mbox{.}}{2013}]%
        {ghosh2013communication}
\bibfield{author}{\bibinfo{person}{Debomita Ghosh}, \bibinfo{person}{T Ghose},
  {and} \bibinfo{person}{Dusmanta~Kumar Mohanta}.}
  \bibinfo{year}{2013}\natexlab{}.
\newblock \showarticletitle{Communication Feasibility Analysis for Smart Grid
  with Phasor Measurement Units}.
\newblock \bibinfo{journal}{{\em {IEEE} Trans. Ind. Informat.\/}}
  \bibinfo{volume}{9}, \bibinfo{number}{3} (\bibinfo{year}{2013}),
  \bibinfo{pages}{1486--1496}.
\newblock


\bibitem[\protect\citeauthoryear{Hoel, Driggs-Campbell, Wolff, Laine, and
  Kochenderfer}{Hoel et~al\mbox{.}}{2019}]%
        {hoel2019combining}
\bibfield{author}{\bibinfo{person}{Carl-Johan Hoel}, \bibinfo{person}{Katherine
  Driggs-Campbell}, \bibinfo{person}{Krister Wolff}, \bibinfo{person}{Leo
  Laine}, {and} \bibinfo{person}{Mykel Kochenderfer}.}
  \bibinfo{year}{2019}\natexlab{}.
\newblock \showarticletitle{Combining Planning and Deep Reinforcement Learning
  in Tactical Decision Making for Autonomous Driving}.
\newblock \bibinfo{journal}{{\em {IEEE} Trans. Intell. Veh.\/}}
  \bibinfo{volume}{9}, \bibinfo{number}{3} (\bibinfo{year}{2019}),
  \bibinfo{pages}{118--122}.
\newblock


\bibitem[\protect\citeauthoryear{Hong, Kandula, Mahajan, Zhang, Gill,
  et~al\mbox{.}}{Hong et~al\mbox{.}}{2013}]%
        {hong2013achieving}
\bibfield{author}{\bibinfo{person}{Chi-Yao Hong}, \bibinfo{person}{Srikanth
  Kandula}, \bibinfo{person}{Ratul Mahajan}, \bibinfo{person}{Ming Zhang},
  \bibinfo{person}{Vijay Gill}, {et~al\mbox{.}}}
  \bibinfo{year}{2013}\natexlab{}.
\newblock \showarticletitle{Achieving High Utilization with Software-Driven
  {WAN}}.
\newblock \bibinfo{journal}{{\em {ACM} {SIGCOMM} {CCR}\/}}
  \bibinfo{volume}{43}, \bibinfo{number}{4} (\bibinfo{year}{2013}),
  \bibinfo{pages}{15--26}.
\newblock


\bibitem[\protect\citeauthoryear{Hong, Mandal, Al-Fares, Zhu, Alimi,
  et~al\mbox{.}}{Hong et~al\mbox{.}}{2018}]%
        {hong2018b4}
\bibfield{author}{\bibinfo{person}{Chi-Yao Hong}, \bibinfo{person}{Subhasree
  Mandal}, \bibinfo{person}{Mohammad Al-Fares}, \bibinfo{person}{Min Zhu},
  \bibinfo{person}{Richard Alimi}, {et~al\mbox{.}}}
  \bibinfo{year}{2018}\natexlab{}.
\newblock \showarticletitle{{B4} and After: Managing Hierarchy, Partitioning,
  and Asymmetry for Availability and Scale in {Google's} Software-Defined
  {WAN}}. In \bibinfo{booktitle}{{\em Proc. {ACM} {SIGCOMM}}}.
\newblock


\bibitem[\protect\citeauthoryear{Hu, Yao, Jin, Zhao, Hu, et~al\mbox{.}}{Hu
  et~al\mbox{.}}{2015}]%
        {hu2015data}
\bibfield{author}{\bibinfo{person}{Shaohan Hu}, \bibinfo{person}{Shuochao Yao},
  \bibinfo{person}{Haiming Jin}, \bibinfo{person}{Yiran Zhao},
  \bibinfo{person}{Yitao Hu}, {et~al\mbox{.}}} \bibinfo{year}{2015}\natexlab{}.
\newblock \showarticletitle{Data Acquisition for Real-Time Decision-Making
  Under Freshness Constraints}. In \bibinfo{booktitle}{{\em Proc. IEEE
  Real-Time Syst. Symp.}} IEEE, \bibinfo{pages}{185--194}.
\newblock


\bibitem[\protect\citeauthoryear{Huang and Modiano}{Huang and Modiano}{2015}]%
        {huang2015optimizing}
\bibfield{author}{\bibinfo{person}{Longbo Huang} {and} \bibinfo{person}{Eytan
  Modiano}.} \bibinfo{year}{2015}\natexlab{}.
\newblock \showarticletitle{Optimizing Age-of-Information in a Multi-Class
  Queueing System}. In \bibinfo{booktitle}{{\em Proc. {IEEE} {ISIT}}}. IEEE,
  \bibinfo{pages}{1681--1685}.
\newblock


\bibitem[\protect\citeauthoryear{Jacobson, Braden, Floyd, Davie, Estrin,
  et~al\mbox{.}}{Jacobson et~al\mbox{.}}{1998}]%
        {jacobson1998recommendations}
\bibfield{author}{\bibinfo{person}{Van Jacobson}, \bibinfo{person}{Bob Braden},
  \bibinfo{person}{Sally Floyd}, \bibinfo{person}{Bruce Davie},
  \bibinfo{person}{Deborah Estrin}, {et~al\mbox{.}}}
  \bibinfo{year}{1998}\natexlab{}.
\newblock \bibinfo{title}{{RFC} 2309: Recommendations on Queue Management and
  Congestion Avoidance in the {Internet}}.
\newblock   (\bibinfo{year}{1998}).
\newblock


\bibitem[\protect\citeauthoryear{Jain, Kumar, Mandal, Ong, Poutievski,
  et~al\mbox{.}}{Jain et~al\mbox{.}}{2013}]%
        {jain2013b4}
\bibfield{author}{\bibinfo{person}{Sushant Jain}, \bibinfo{person}{Alok Kumar},
  \bibinfo{person}{Subhasree Mandal}, \bibinfo{person}{Joon Ong},
  \bibinfo{person}{Leon Poutievski}, {et~al\mbox{.}}}
  \bibinfo{year}{2013}\natexlab{}.
\newblock \showarticletitle{{B4}: Experience with a Globally-Deployed Software
  Defined {WAN}}.
\newblock \bibinfo{journal}{{\em {ACM} {SIGCOMM} {CCR}\/}}
  \bibinfo{volume}{43}, \bibinfo{number}{4} (\bibinfo{year}{2013}),
  \bibinfo{pages}{3--14}.
\newblock


\bibitem[\protect\citeauthoryear{Kadota, Rahman, and Modiano}{Kadota
  et~al\mbox{.}}{2020}]%
        {kadota2020age}
\bibfield{author}{\bibinfo{person}{Igor Kadota}, \bibinfo{person}{M~Shahir
  Rahman}, {and} \bibinfo{person}{Eytan Modiano}.}
  \bibinfo{year}{2020}\natexlab{}.
\newblock \showarticletitle{Age of Information in Wireless Networks: From
  Theory to Implementation}. In \bibinfo{booktitle}{{\em Proc. {IEEE}
  {MobiCom}}}. \bibinfo{pages}{1--3}.
\newblock


\bibitem[\protect\citeauthoryear{Kadota, Sinha, and Modiano}{Kadota
  et~al\mbox{.}}{2018a}]%
        {kadota2018optimizing}
\bibfield{author}{\bibinfo{person}{Igor Kadota}, \bibinfo{person}{Abhishek
  Sinha}, {and} \bibinfo{person}{Eytan Modiano}.}
  \bibinfo{year}{2018}\natexlab{a}.
\newblock \showarticletitle{Optimizing Age of Information in Wireless Networks
  with Throughput Constraints}. In \bibinfo{booktitle}{{\em Proc. {IEEE}
  {INFOCOM}}}. IEEE, \bibinfo{pages}{1844--1852}.
\newblock


\bibitem[\protect\citeauthoryear{Kadota, Sinha, and Modiano}{Kadota
  et~al\mbox{.}}{2019}]%
        {kadota2019scheduling}
\bibfield{author}{\bibinfo{person}{Igor Kadota}, \bibinfo{person}{Abhishek
  Sinha}, {and} \bibinfo{person}{Eytan Modiano}.}
  \bibinfo{year}{2019}\natexlab{}.
\newblock \showarticletitle{Scheduling Algorithms for Optimizing Age of
  Information in Wireless Networks with Throughput Constraints}.
\newblock \bibinfo{journal}{{\em {IEEE/ACM} Trans. Netw.\/}}
  \bibinfo{volume}{27}, \bibinfo{number}{4} (\bibinfo{year}{2019}),
  \bibinfo{pages}{1359--1372}.
\newblock


\bibitem[\protect\citeauthoryear{Kadota, Sinha, Uysal-Biyikoglu, Singh, and
  Modiano}{Kadota et~al\mbox{.}}{2018b}]%
        {kadota2018scheduling}
\bibfield{author}{\bibinfo{person}{Igor Kadota}, \bibinfo{person}{Abhishek
  Sinha}, \bibinfo{person}{Elif Uysal-Biyikoglu}, \bibinfo{person}{Rahul
  Singh}, {and} \bibinfo{person}{Eytan Modiano}.}
  \bibinfo{year}{2018}\natexlab{b}.
\newblock \showarticletitle{Scheduling Policies for Minimizing Age of
  Information in Broadcast Wireless Networks}.
\newblock \bibinfo{journal}{{\em {IEEE/ACM} Trans. Netw.\/}}
  \bibinfo{volume}{26}, \bibinfo{number}{6} (\bibinfo{year}{2018}),
  \bibinfo{pages}{2637--2650}.
\newblock


\bibitem[\protect\citeauthoryear{Kam, Kompella, and Ephremides}{Kam
  et~al\mbox{.}}{2013}]%
        {kam2013age}
\bibfield{author}{\bibinfo{person}{Clement Kam}, \bibinfo{person}{Sastry
  Kompella}, {and} \bibinfo{person}{Anthony Ephremides}.}
  \bibinfo{year}{2013}\natexlab{}.
\newblock \showarticletitle{Age of Information Under Random Updates}. In
  \bibinfo{booktitle}{{\em Proc. {IEEE} {ISIT}}}. IEEE,
  \bibinfo{pages}{66--70}.
\newblock


\bibitem[\protect\citeauthoryear{Kang, Son, and Stankovic}{Kang
  et~al\mbox{.}}{2004}]%
        {kang2004managing}
\bibfield{author}{\bibinfo{person}{K-D Kang}, \bibinfo{person}{Sang~Hyuk Son},
  {and} \bibinfo{person}{John~A Stankovic}.} \bibinfo{year}{2004}\natexlab{}.
\newblock \showarticletitle{Managing Deadline Miss Ratio and Sensor Data
  Freshness in Real-Time Databases}.
\newblock \bibinfo{journal}{{\em {IEEE} Trans. Knowl. Data Eng.\/}}
  \bibinfo{volume}{16}, \bibinfo{number}{10} (\bibinfo{year}{2004}),
  \bibinfo{pages}{1200--1216}.
\newblock


\bibitem[\protect\citeauthoryear{Kaul, Gruteser, Rai, and Kenney}{Kaul
  et~al\mbox{.}}{2011}]%
        {kaul2011minimizing}
\bibfield{author}{\bibinfo{person}{Sanjit Kaul}, \bibinfo{person}{Marco
  Gruteser}, \bibinfo{person}{Vinuth Rai}, {and} \bibinfo{person}{John
  Kenney}.} \bibinfo{year}{2011}\natexlab{}.
\newblock \showarticletitle{Minimizing Age of Information in Vehicular
  Networks}. In \bibinfo{booktitle}{{\em Proc. {IEEE} {SECON}}}. IEEE,
  \bibinfo{pages}{350--358}.
\newblock


\bibitem[\protect\citeauthoryear{Kaul, Yates, and Gruteser}{Kaul
  et~al\mbox{.}}{2012}]%
        {kaul2012real}
\bibfield{author}{\bibinfo{person}{Sanjit Kaul}, \bibinfo{person}{Roy Yates},
  {and} \bibinfo{person}{Marco Gruteser}.} \bibinfo{year}{2012}\natexlab{}.
\newblock \showarticletitle{Real-Time Status: How Often Should One Update?}. In
  \bibinfo{booktitle}{{\em Proc. {IEEE} {INFOCOM}}}. IEEE,
  \bibinfo{pages}{2731--2735}.
\newblock


\bibitem[\protect\citeauthoryear{Kavitha, Altman, and Saha}{Kavitha
  et~al\mbox{.}}{2018}]%
        {kavitha2018controlling}
\bibfield{author}{\bibinfo{person}{Veeraruna Kavitha}, \bibinfo{person}{Eitan
  Altman}, {and} \bibinfo{person}{Indrajit Saha}.}
  \bibinfo{year}{2018}\natexlab{}.
\newblock \showarticletitle{Controlling Packet Drops to Improve Freshness of
  Information}.
\newblock \bibinfo{journal}{{\em arXiv preprint arXiv:1807.09325\/}}
  (\bibinfo{year}{2018}).
\newblock


\bibitem[\protect\citeauthoryear{Kim, Abdelzaher, Sha, Bar-Noy, and Hobbs}{Kim
  et~al\mbox{.}}{2016}]%
        {kim2016sporadic}
\bibfield{author}{\bibinfo{person}{Jung-Eun Kim}, \bibinfo{person}{Tarek
  Abdelzaher}, \bibinfo{person}{Lui Sha}, \bibinfo{person}{Amotz Bar-Noy},
  {and} \bibinfo{person}{Reginald Hobbs}.} \bibinfo{year}{2016}\natexlab{}.
\newblock \showarticletitle{Sporadic Decision-Centric Data Scheduling with
  Normally-Off Sensors}. In \bibinfo{booktitle}{{\em 2016 IEEE Real-Time
  Systems Symposium (RTSS)}}. IEEE, \bibinfo{pages}{135--145}.
\newblock


\bibitem[\protect\citeauthoryear{Laoutaris, Sirivianos, Yang, and
  Rodriguez}{Laoutaris et~al\mbox{.}}{2011}]%
        {laoutaris2011inter}
\bibfield{author}{\bibinfo{person}{Nikolaos Laoutaris},
  \bibinfo{person}{Michael Sirivianos}, \bibinfo{person}{Xiaoyuan Yang}, {and}
  \bibinfo{person}{Pablo Rodriguez}.} \bibinfo{year}{2011}\natexlab{}.
\newblock \showarticletitle{Inter-Datacenter Bulk Transfers with
  {NetStitcher}}.
\newblock \bibinfo{journal}{{\em {ACM} {SIGCOMM} {CCR}\/}}
  \bibinfo{volume}{41}, \bibinfo{number}{4} (\bibinfo{year}{2011}),
  \bibinfo{pages}{74--85}.
\newblock


\bibitem[\protect\citeauthoryear{{Lawton}, {Young}, and {Beard}}{{Lawton}
  et~al\mbox{.}}{2000}]%
        {lawton00}
\bibfield{author}{\bibinfo{person}{A.~R. {Lawton}}, \bibinfo{person}{B.~J.
  {Young}}, {and} \bibinfo{person}{R.~W. {Beard}}.}
  \bibinfo{year}{2000}\natexlab{}.
\newblock \showarticletitle{A decentralized approach to elementary formation
  manoeuvres}. In \bibinfo{booktitle}{{\em Proc. {IEEE} ICRA}},
  Vol.~\bibinfo{volume}{3}. \bibinfo{pages}{2728--2733}.
\newblock


\bibitem[\protect\citeauthoryear{Liu, Wang, Bai, and Dai}{Liu
  et~al\mbox{.}}{2018}]%
        {liu2018age}
\bibfield{author}{\bibinfo{person}{Juan Liu}, \bibinfo{person}{Xijun Wang},
  \bibinfo{person}{Bo Bai}, {and} \bibinfo{person}{Huaiyu Dai}.}
  \bibinfo{year}{2018}\natexlab{}.
\newblock \showarticletitle{Age-Optimal Trajectory Planning for {UAV-Assisted}
  Data Collection}. In \bibinfo{booktitle}{{\em IEEE INFOCOM 2018 Workshops
  (INFOCOM WKSHPS)}}. IEEE, \bibinfo{pages}{553--558}.
\newblock


\bibitem[\protect\citeauthoryear{Lou, Yuan, Kompella, and Tzeng}{Lou
  et~al\mbox{.}}{2020}]%
        {lou2020aoi}
\bibfield{author}{\bibinfo{person}{Jiadong Lou}, \bibinfo{person}{Xu Yuan},
  \bibinfo{person}{Sastry Kompella}, {and} \bibinfo{person}{Nian-Feng Tzeng}.}
  \bibinfo{year}{2020}\natexlab{}.
\newblock \showarticletitle{{AoI} and Throughput Tradeoffs in Routing-Aware
  Multi-Hop Wireless Networks}. In \bibinfo{booktitle}{{\em Proc. {IEEE}
  {INFOCOM}}}. IEEE, \bibinfo{pages}{476--485}.
\newblock


\bibitem[\protect\citeauthoryear{Lou, Yuan, Kompella, and Tzeng}{Lou
  et~al\mbox{.}}{2021}]%
        {lou2021boosting}
\bibfield{author}{\bibinfo{person}{Jiadong Lou}, \bibinfo{person}{Xu Yuan},
  \bibinfo{person}{Sastry Kompella}, {and} \bibinfo{person}{Nian-Feng Tzeng}.}
  \bibinfo{year}{2021}\natexlab{}.
\newblock \showarticletitle{Boosting or Hindering: {AoI} and Throughput
  Interrelation in Routing-Aware Multi-Hop Wireless Networks}.
\newblock \bibinfo{journal}{{\em {IEEE/ACM} Trans. Netw.\/}}
  (\bibinfo{year}{2021}).
\newblock


\bibitem[\protect\citeauthoryear{Lu, Ji, and Li}{Lu et~al\mbox{.}}{2018}]%
        {lu2018age}
\bibfield{author}{\bibinfo{person}{Ning Lu}, \bibinfo{person}{Bo Ji}, {and}
  \bibinfo{person}{Bin Li}.} \bibinfo{year}{2018}\natexlab{}.
\newblock \showarticletitle{Age-Based Scheduling: Improving Data Freshness for
  Wireless Real-Time Traffic}. In \bibinfo{booktitle}{{\em Proc. {ACM}
  {Mobihoc}}}. \bibinfo{pages}{191--200}.
\newblock


\bibitem[\protect\citeauthoryear{Naduvathuparambil, Valenti, and
  Feliachi}{Naduvathuparambil et~al\mbox{.}}{2002}]%
        {naduvathuparambil2002communication}
\bibfield{author}{\bibinfo{person}{Biju Naduvathuparambil},
  \bibinfo{person}{Matthew~C Valenti}, {and} \bibinfo{person}{Ali Feliachi}.}
  \bibinfo{year}{2002}\natexlab{}.
\newblock \showarticletitle{Communication Delays in Wide Area Measurement
  Systems}. In \bibinfo{booktitle}{{\em Proc. 34th Southeastern Symp. Syst.
  Theory}}. IEEE, \bibinfo{pages}{118--122}.
\newblock


\bibitem[\protect\citeauthoryear{Nichols and Jacobson}{Nichols and
  Jacobson}{2012}]%
        {nichols2012controlling}
\bibfield{author}{\bibinfo{person}{Kathleen Nichols} {and} \bibinfo{person}{Van
  Jacobson}.} \bibinfo{year}{2012}\natexlab{}.
\newblock \showarticletitle{Controlling Queue Delay}.
\newblock \bibinfo{journal}{{\it Commun. ACM}} \bibinfo{volume}{55},
  \bibinfo{number}{7} (\bibinfo{year}{2012}), \bibinfo{pages}{42--50}.
\newblock


\bibitem[\protect\citeauthoryear{Pan, Natarajan, Piglione, Prabhu, Subramanian,
  et~al\mbox{.}}{Pan et~al\mbox{.}}{2013}]%
        {pan2013pie}
\bibfield{author}{\bibinfo{person}{Rong Pan}, \bibinfo{person}{Preethi
  Natarajan}, \bibinfo{person}{Chiara Piglione},
  \bibinfo{person}{Mythili~Suryanarayana Prabhu}, \bibinfo{person}{Vijay
  Subramanian}, {et~al\mbox{.}}} \bibinfo{year}{2013}\natexlab{}.
\newblock \showarticletitle{{PIE}: A Lightweight Control Scheme to Address the
  Bufferbloat Problem}. In \bibinfo{booktitle}{{\em Proc. {IEEE} {HPSR}}}.
  \bibinfo{pages}{148--155}.
\newblock


\bibitem[\protect\citeauthoryear{Pappas, Gunnarsson, Kratz, Kountouris, and
  Angelakis}{Pappas et~al\mbox{.}}{2015}]%
        {pappas2015age}
\bibfield{author}{\bibinfo{person}{Nikolaos Pappas}, \bibinfo{person}{Johan
  Gunnarsson}, \bibinfo{person}{Ludvig Kratz}, \bibinfo{person}{Marios
  Kountouris}, {and} \bibinfo{person}{Vangelis Angelakis}.}
  \bibinfo{year}{2015}\natexlab{}.
\newblock \showarticletitle{Age of Information of Multiple Sources with Queue
  Management}. In \bibinfo{booktitle}{{\em Proc. {IEEE} {ICC}}}. IEEE,
  \bibinfo{pages}{5935--5940}.
\newblock


\bibitem[\protect\citeauthoryear{Phadke}{Phadke}{1993}]%
        {phadke1993synchronized}
\bibfield{author}{\bibinfo{person}{Arun~G Phadke}.}
  \bibinfo{year}{1993}\natexlab{}.
\newblock \showarticletitle{Synchronized Phasor Measurements in Power Systems}.
\newblock \bibinfo{journal}{{\em IEEE Computer Applications in power\/}}
  \bibinfo{volume}{6}, \bibinfo{number}{2} (\bibinfo{year}{1993}),
  \bibinfo{pages}{10--15}.
\newblock


\bibitem[\protect\citeauthoryear{Primadianto and Lu}{Primadianto and
  Lu}{2016}]%
        {primadianto2016review}
\bibfield{author}{\bibinfo{person}{Anggoro Primadianto} {and}
  \bibinfo{person}{Chan-Nan Lu}.} \bibinfo{year}{2016}\natexlab{}.
\newblock \showarticletitle{A Review on Distribution System State Estimation}.
\newblock \bibinfo{journal}{{\em {IEEE} Trans. Power Syst.\/}}
  \bibinfo{volume}{32}, \bibinfo{number}{5} (\bibinfo{year}{2016}),
  \bibinfo{pages}{3875--3883}.
\newblock


\bibitem[\protect\citeauthoryear{Shreedhar, Kaul, and Yates}{Shreedhar
  et~al\mbox{.}}{2018}]%
        {shreedhar2018acp}
\bibfield{author}{\bibinfo{person}{Tanya Shreedhar}, \bibinfo{person}{Sanjit~K
  Kaul}, {and} \bibinfo{person}{Roy~D Yates}.} \bibinfo{year}{2018}\natexlab{}.
\newblock \showarticletitle{{ACP}: Age Control Protocol for Minimizing Age of
  Information Over the {Internet}}. In \bibinfo{booktitle}{{\em Proc. {IEEE}
  {MobiCom}}}. \bibinfo{pages}{699--701}.
\newblock


\bibitem[\protect\citeauthoryear{Sun, Uysal-Biyikoglu, Yates, Koksal, and
  Shroff}{Sun et~al\mbox{.}}{2016}]%
        {sun2016update}
\bibfield{author}{\bibinfo{person}{Yin Sun}, \bibinfo{person}{Elif
  Uysal-Biyikoglu}, \bibinfo{person}{Roy Yates}, \bibinfo{person}{C~Emre
  Koksal}, {and} \bibinfo{person}{Ness~B Shroff}.}
  \bibinfo{year}{2016}\natexlab{}.
\newblock \showarticletitle{Update or Wait: How to Keep Your Data Fresh}. In
  \bibinfo{booktitle}{{\em Proc. {IEEE} {INFOCOM}}}. IEEE,
  \bibinfo{pages}{1--9}.
\newblock


\bibitem[\protect\citeauthoryear{Sun, Uysal-Biyikoglu, Yates, Koksal, and
  Shroff}{Sun et~al\mbox{.}}{2017}]%
        {sun2017update}
\bibfield{author}{\bibinfo{person}{Yin Sun}, \bibinfo{person}{Elif
  Uysal-Biyikoglu}, \bibinfo{person}{Roy~D Yates}, \bibinfo{person}{C~Emre
  Koksal}, {and} \bibinfo{person}{Ness~B Shroff}.}
  \bibinfo{year}{2017}\natexlab{}.
\newblock \showarticletitle{Update or Wait: How to Keep Your Data Fresh}.
\newblock \bibinfo{journal}{{\em {IEEE} Trans. Inf. Theory\/}}
  \bibinfo{volume}{63}, \bibinfo{number}{11} (\bibinfo{year}{2017}),
  \bibinfo{pages}{7492--7508}.
\newblock


\bibitem[\protect\citeauthoryear{Taht, Gettys, Dumazet, Hoeiland-Joergensen,
  Dumazet, et~al\mbox{.}}{Taht et~al\mbox{.}}{2018}]%
        {taht2018flow}
\bibfield{author}{\bibinfo{person}{Dave Taht}, \bibinfo{person}{Jim Gettys},
  \bibinfo{person}{E Dumazet}, \bibinfo{person}{Toke Hoeiland-Joergensen},
  \bibinfo{person}{Eric Dumazet}, {et~al\mbox{.}}}
  \bibinfo{year}{2018}\natexlab{}.
\newblock \bibinfo{title}{{RFC} 8290: The Flow Queue CoDel Packet Scheduler and
  Active Queue Management Algorithm}.
\newblock   (\bibinfo{year}{2018}).
\newblock


\bibitem[\protect\citeauthoryear{Talak, Kadota, Karaman, and Modiano}{Talak
  et~al\mbox{.}}{2018}]%
        {talak2018scheduling}
\bibfield{author}{\bibinfo{person}{Rajat Talak}, \bibinfo{person}{Igor Kadota},
  \bibinfo{person}{Sertac Karaman}, {and} \bibinfo{person}{Eytan Modiano}.}
  \bibinfo{year}{2018}\natexlab{}.
\newblock \showarticletitle{Scheduling Policies for Age Minimization in
  Wireless Networks with Unknown Channel State}. In \bibinfo{booktitle}{{\em
  Proc. {IEEE} {ISIT}}}. IEEE, \bibinfo{pages}{2564--2568}.
\newblock


\bibitem[\protect\citeauthoryear{Talak, Karaman, and Modiano}{Talak
  et~al\mbox{.}}{2016}]%
        {talak2016speed}
\bibfield{author}{\bibinfo{person}{Rajat Talak}, \bibinfo{person}{Sertac
  Karaman}, {and} \bibinfo{person}{Eytan Modiano}.}
  \bibinfo{year}{2016}\natexlab{}.
\newblock \showarticletitle{Speed Limits in Autonomous Vehicular Networks Due
  to Communication Constraints}. In \bibinfo{booktitle}{{\em Proc. {IEEE}
  {CDC}}}. IEEE, \bibinfo{pages}{4998--5003}.
\newblock


\bibitem[\protect\citeauthoryear{Talak, Karaman, and Modiano}{Talak
  et~al\mbox{.}}{2019}]%
        {talak2019optimizing}
\bibfield{author}{\bibinfo{person}{Rajat Talak}, \bibinfo{person}{Sertac
  Karaman}, {and} \bibinfo{person}{Eytan Modiano}.}
  \bibinfo{year}{2019}\natexlab{}.
\newblock \showarticletitle{Optimizing Information Freshness in Wireless
  Networks Under General Interference Constraints}.
\newblock \bibinfo{journal}{{\em {IEEE/ACM} Trans. Netw.\/}}
  (\bibinfo{year}{2019}).
\newblock


\bibitem[\protect\citeauthoryear{Talak, Karaman, and Modiano}{Talak
  et~al\mbox{.}}{2020}]%
        {talak2020improving}
\bibfield{author}{\bibinfo{person}{Rajat Talak}, \bibinfo{person}{Sertac
  Karaman}, {and} \bibinfo{person}{Eytan Modiano}.}
  \bibinfo{year}{2020}\natexlab{}.
\newblock \showarticletitle{Improving Age of Information in Wireless Networks
  with Perfect Channel State Information}.
\newblock \bibinfo{journal}{{\em {IEEE/ACM} Trans. Netw.\/}}
  \bibinfo{volume}{28}, \bibinfo{number}{4} (\bibinfo{year}{2020}),
  \bibinfo{pages}{1765--1778}.
\newblock


\bibitem[\protect\citeauthoryear{Tardioli, Sicignano, Riazuelo, Romeo,
  Villarroel, et~al\mbox{.}}{Tardioli et~al\mbox{.}}{2016}]%
        {tardioli16}
\bibfield{author}{\bibinfo{person}{Danilo Tardioli}, \bibinfo{person}{Domenico
  Sicignano}, \bibinfo{person}{Luis Riazuelo}, \bibinfo{person}{Antonio Romeo},
  \bibinfo{person}{Jos{\'e}~Luis Villarroel}, {et~al\mbox{.}}}
  \bibinfo{year}{2016}\natexlab{}.
\newblock \showarticletitle{Robot Teams for Intervention in Confined and
  Structured Environments}.
\newblock \bibinfo{journal}{{\em J. Field Robotics\/}}  \bibinfo{volume}{33}
  (\bibinfo{year}{2016}), \bibinfo{pages}{765--801}.
\newblock


\bibitem[\protect\citeauthoryear{Terzija, Valverde, Cai, Regulski, Madani,
  et~al\mbox{.}}{Terzija et~al\mbox{.}}{2011}]%
        {terzija2011wide}
\bibfield{author}{\bibinfo{person}{Vladimir Terzija}, \bibinfo{person}{Gustavo
  Valverde}, \bibinfo{person}{Deyu Cai}, \bibinfo{person}{Pawel Regulski},
  \bibinfo{person}{Vahid Madani}, {et~al\mbox{.}}}
  \bibinfo{year}{2011}\natexlab{}.
\newblock \showarticletitle{Wide-Area Monitoring, Protection, and Control of
  Future Electric Power Networks}.
\newblock \bibinfo{journal}{{\it Proc. IEEE}} \bibinfo{volume}{99},
  \bibinfo{number}{1} (\bibinfo{year}{2011}), \bibinfo{pages}{80--93}.
\newblock


\bibitem[\protect\citeauthoryear{Tripathi, Talak, and Modiano}{Tripathi
  et~al\mbox{.}}{2019}]%
        {tripathi2019age}
\bibfield{author}{\bibinfo{person}{Vishrant Tripathi}, \bibinfo{person}{Rajat
  Talak}, {and} \bibinfo{person}{Eytan Modiano}.}
  \bibinfo{year}{2019}\natexlab{}.
\newblock \showarticletitle{Age Optimal Information Gathering and Dissemination
  on Graphs}. In \bibinfo{booktitle}{{\em Proc. {IEEE} {INFOCOM}}}. IEEE,
  \bibinfo{pages}{2422--2430}.
\newblock


\bibitem[\protect\citeauthoryear{Tripathi, Talak, and Modiano}{Tripathi
  et~al\mbox{.}}{2021}]%
        {tripathi2021age}
\bibfield{author}{\bibinfo{person}{Vishrant Tripathi}, \bibinfo{person}{Rajat
  Talak}, {and} \bibinfo{person}{Eytan Modiano}.}
  \bibinfo{year}{2021}\natexlab{}.
\newblock \showarticletitle{Age Optimal Information Gathering and Dissemination
  on Graphs}.
\newblock \bibinfo{journal}{{\em {IEEE} Trans. Mobile Comput.\/}}
  (\bibinfo{year}{2021}).
\newblock


\bibitem[\protect\citeauthoryear{Wilson, Ballani, Karagiannis, and
  Rowtron}{Wilson et~al\mbox{.}}{2011}]%
        {wilson2011better}
\bibfield{author}{\bibinfo{person}{Christo Wilson}, \bibinfo{person}{Hitesh
  Ballani}, \bibinfo{person}{Thomas Karagiannis}, {and} \bibinfo{person}{Ant
  Rowtron}.} \bibinfo{year}{2011}\natexlab{}.
\newblock \showarticletitle{Better Never Than Late: Meeting Deadlines in
  Datacenter Networks}.
\newblock \bibinfo{journal}{{\em {ACM} {SIGCOMM} {CCR}\/}}
  \bibinfo{volume}{41}, \bibinfo{number}{4} (\bibinfo{year}{2011}),
  \bibinfo{pages}{50--61}.
\newblock


\bibitem[\protect\citeauthoryear{Yates}{Yates}{2015}]%
        {yates2015lazy}
\bibfield{author}{\bibinfo{person}{Roy~D Yates}.}
  \bibinfo{year}{2015}\natexlab{}.
\newblock \showarticletitle{Lazy is Timely: Status Updates by an Energy
  Harvesting Source}. In \bibinfo{booktitle}{{\em Proc. {IEEE} {ISIT}}}. IEEE,
  \bibinfo{pages}{3008--3012}.
\newblock


\bibitem[\protect\citeauthoryear{Yates and Kaul}{Yates and Kaul}{2012}]%
        {yates2012real}
\bibfield{author}{\bibinfo{person}{Roy~D Yates} {and} \bibinfo{person}{Sanjit
  Kaul}.} \bibinfo{year}{2012}\natexlab{}.
\newblock \showarticletitle{Real-Time Status Updating: Multiple Sources}. In
  \bibinfo{booktitle}{{\em Proc. {IEEE} {ISIT}}}. IEEE,
  \bibinfo{pages}{2666--2670}.
\newblock


\bibitem[\protect\citeauthoryear{Yates and Kaul}{Yates and Kaul}{2019}]%
        {yates2019age}
\bibfield{author}{\bibinfo{person}{Roy~D Yates} {and} \bibinfo{person}{Sanjit~K
  Kaul}.} \bibinfo{year}{2019}\natexlab{}.
\newblock \showarticletitle{The Age of Information: Real-Time Status Updating
  by Multiple Sources}.
\newblock \bibinfo{journal}{{\em {IEEE} Trans. Inf. Theory\/}}
  \bibinfo{volume}{65}, \bibinfo{number}{3} (\bibinfo{year}{2019}),
  \bibinfo{pages}{1807--1827}.
\newblock


\end{thebibliography}

\appendix
\section{NP-Hardness to Determine a Zero-Queueing-Delay Schedule}
We prove \thm{zero-queueing-delay} by reducing 3-SAT to it. Since 3-SAT is NP-complete, we conclude that obtaining a zero-queueing-delay scheduling is NP-hard. Below we construct such a reduction.

We express a 3-SAT instance $\phi$ in the form of
\begin{align*}
\phi = C_1 \wedge C_2 \wedge \cdots \wedge C_n
\end{align*}
where $C_k = y_{k,1} \vee y_{k,2} \vee y_{k,3}$ is a clause with literals $y_{k,1}, y_{k,2},$ and $y_{k,3}$. Each literal $y_{k,j}$ belongs to either the given set $Y$ of literals or its complement $\overline{Y}$.

\begin{figure}
\centering
\includegraphics[scale=1]{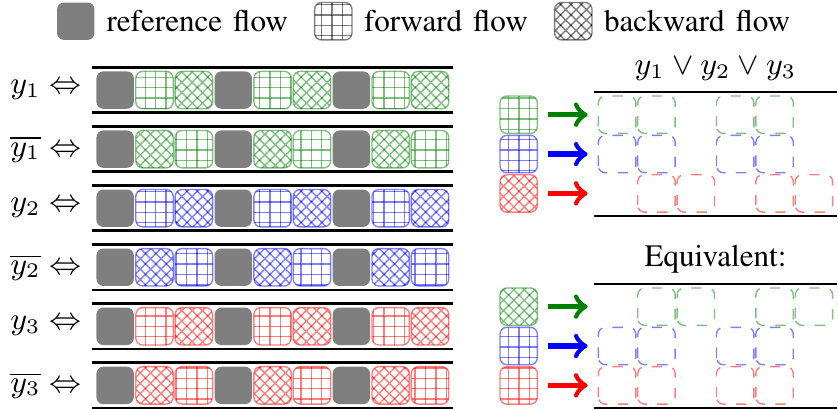}
\caption{We show that a zero-queueing-delay schedule is NP-hard to obtain by reducing 3-SAT, an NP-complete problem, to it. For each literal $y_k$ in the 3-SAT instance, we create a unit-capacity throttling link and two AoI flows, the forward and the backward flow, through it. We send a reference flow through all throttling links. Without queueing, there are two possible ways the flows could share the throttling link, and we associate the two configurations to $y_k$ and $\overline{y_k}$. Leveraging the forward and backward flows, we create a link with capacity $2$ for each clause such that a zero-queueing-delay schedule would not exceed the link capacity. As such, a satisfiable 3-SAT instance corresponds to a zero-queueing-delay schedule.}
\label{fig:proof_3SAT}
\end{figure}

As shown in \fig{proof_3SAT}, for each literal $y \in Y$, we construct one unit-capacity throttling link for two AoI flows, the forward flow and the backward flow, with update frequency $\frac{1}{3}$ and size $1$. 
We assign a reference AoI flow through all these throttling links so that there are two possible zero-queueing-delay configurations. We associate the configurations to $y$ and $\overline{y}$ respectively. 
Each clause corresponds to a link of capacity $2$, each carries $3$ AoI flows according to the literals. Two of them are forward flows and one is backward flow. We insert propagation delay appropriately so that the clause is satisfied if and only if the link capacity is not exceeded.
There are three equivalent ways to choose the forward/backward flows, and we simply choose one of them.

In this way, all the clauses are satisfiable if and only if the queueing delay is zero. Therefore, a zero-queueing-delay schedule is equivalent to a feasible solution to 3-SAT, which concludes the proof.

\section{AoI Upper Bound}
Below we prove \thm{bounded-queueing-delay} by constructing a scheduling policy with the desired upper-bounded $a_f$. We first assume all link latency is zero. For each $f \in F_\AoI$ with $\mu_f > 0$, we allocate capacity $\frac{c_l \mu_f s_f}{S_\LDA + S_\AoI}$ to send the flow. As such, since a feasible solution to LAC satisfies \eqn{cons:capacity}, each packet is sent within
\begin{align*}
\frac{s_f (S_\LDA + S_\AoI)}{c_l \mu_f s_f} \leq \frac{s_f}{\mu_f s_f} = \frac{1}{\mu_f}
\end{align*}
upon arrival. In other words, it takes at most $\frac{1}{\mu_f}$ to go through each link on the path of $f$. Therefore, we have
\begin{align*}
a_f \leq \frac{1}{2 \mu_f} + |f| \cdot \frac{1}{\mu_f} = \frac{1 + 2|f|}{2 \mu_f}.
\end{align*}
For non-zero link latency, we add $d_f$ to the right hand side of the inequality, which leads to the desired result.

\end{document}